\documentclass[preprint,prd,tightenlines,nofootinbib,eqsecnum,superscriptaddress]{revtex4-2}

\pdfoutput=1

\usepackage{amsmath}    
\usepackage{amsfonts}   
\usepackage{amssymb}    
\usepackage{mathtools}  
\usepackage{physics}

\usepackage[T1]{fontenc}   
\usepackage{mathrsfs}      

\DeclareSymbolFontAlphabet{\mathrsfs}{rsfs}
\usepackage{parskip}

\usepackage{graphicx}   

\usepackage{array}               
\usepackage{booktabs}            
\usepackage[table]{xcolor}       

\usepackage[most]{tcolorbox}          
\usepackage{orcidlink}  
\usepackage{hyperref}
\hypersetup{
    colorlinks = true,
    linktocpage,
    unicode,
    linkcolor  = blue,
    filecolor  = magenta,
    urlcolor   = cyan,
}

\definecolor{CiteColor}{rgb}{0, 0.5, 0}       
\definecolor{RefColor} {rgb}{0.55, 0, 0}      
\definecolor{URLColor} {rgb}{0, 0, 0.35}      

\hypersetup{citecolor=CiteColor, linkcolor=RefColor, urlcolor=URLColor}

\newcolumntype{C}{>{$}c<{$}}
\AtBeginDocument{%
  \heavyrulewidth  = .08em
  \lightrulewidth  = .05em
  \cmidrulewidth   = .03em
  \belowrulesep    = .65ex
  \belowbottomsep  = 0pt
  \aboverulesep    = .4ex
  \abovetopsep     = 0pt
  \cmidrulesep     = \doublerulesep
  \cmidrulekern    = .5em
  \defaultaddspace = .5em
}

\newcommand{\quand}{\quad\text{and}\quad}   
\newcommand{\ud}   {\mathrm{d}}             
\newcommand{\ui}   {\mathrm{i}}             
\newcommand{\rl}[1]{\mathrm{Re}\left[#1\right]}
\newcommand{\img}[1]{\mathrm{Im}\left[#1\right]}
\newcommand{\tsf}[1]{\textsf{#1}}
\newcommand{\ostime}{\zeta}

\newtcolorbox{mybox}{
  colback=blue!5!white,
  colframe=blue!50!black,
  before upper={\selectfont}
}
\newcommand{\GC}{Q}
\newcommand{\shift}{\sigma}
\newcommand{\pt}{\tau}

\newcommand{\mc}{\mathcal}
\newcommand{\mf}{\mathfrak}
\newcommand{\fr}{\frac}          

\newcommand{\ph}       {\phantom{o}}          

\begin{document}

\title{Separable equations of motion for spinning test bodies in spacetimes 
with Killing--Yano symmetry}

\author{Abhinove N.~Seenivasan, \orcidlink{0000-0003-2987-5399}}\,\email{anseenivasan1@sheffield.ac.uk}
\affiliation{Consortium for Fundamental Physics, School of Mathematical and Physical Sciences,
University of Sheffield, Hicks Building, Hounsfield Rd, Sheffield S3 7RH, United Kingdom.}

\author{Paul Ramond, \orcidlink{0000-0001-7123-0039}}\,\email{ramond@lpccaen.in2p3.fr}
\affiliation{Universit\'{e} de Caen Normandie, ENSICAEN, CNRS/IN2P3, LPC Caen UMR6534, F-14000 Caen, France}

\author{Soichiro Isoyama, \orcidlink{0000-0001-6247-2642}}\,\email{isoyama@yukawa.kyoto-u.ac.jp}
\affiliation{Department of Physics, National University of Singapore, 21 Lower Kent Ridge Rd, 119077 Singapore}

\author{Sam R.~Dolan, \orcidlink{0000-0002-4672-6523}}\,\email{s.dolan@sheffield.ac.uk}
\affiliation{Consortium for Fundamental Physics, School of Mathematical and Physical Sciences,
University of Sheffield, Hicks Building, Hounsfield Rd, Sheffield S3 7RH, United Kingdom.}

\date{\today}

\begin{abstract}
At linear order in spin, the motion of a spinning compact body in curved spacetime is known to have two remarkable properties: (i) in Kerr spacetime, a suitable shift of the worldline maps the motion onto fiducial geodesics, yielding separable equations of motion; and (ii) in four-dimensional spacetimes admitting a non-degenerate Killing--Yano tensor, the dynamics are integrable.
In this work, we generalize the worldline-shift method of (i) to the whole class of spacetimes in (ii), using a covariant construction built on the Killing--Yano symmetry. We show that spin then plays the role of charge: after the shift, the motion is described by a fictitious spinless particle, with charge set by the spin component along the orbital angular momentum, obeying a Lorentz-force equation. The effective field strength is built from the Ricci tensor and the dual of the Killing--Yano tensor. It vanishes in Ricci-flat spacetimes, recovering the geodesic description known in Kerr.
In the general off-shell metric admitting a Killing--Yano tensor, we derive explicitly separated equations of motion, which reduce to quadratures and provide a constructive realization of the known integrability. Since no field equations are assumed, our results apply to any such spacetime, including those sourced by matter or arising in modified theories of gravity, and we illustrate them in Kerr--de Sitter spacetime.

\end{abstract}

\maketitle
{\setlength{\parskip}{0pt}\tableofcontents}

\clearpage
\section{Introduction and summary}
\label{sec:intro}

In the 18th and 19th centuries, the study of the motion of spinning bodies under gravity drove the development of the mathematical theory of integrable systems. Indeed, canonical examples of super-integrable, integrable and non-integrable/chaotic systems, respectively, are found in the Kepler two-body problem, the symmetric spinning top (of Euler, Lagrange and Kovalevskaya), and the asymmetric spinning top. A dynamical system is integrable if there exists a sufficient number of independent constants of motion that it becomes solvable by quadratures; and each constant of motion is associated with an underlying symmetry of the dynamical system \cite{Cariglia:2014ysa}.

In the 21st century, experimental advances in gravitational-wave astronomy give a compelling motivation for the study of the dynamics of spinning bodies in curved spacetimes. The GWTC catalogue \cite{GWTC5.26} now comprises more than 300 merging binary systems across two orders of magnitude in mass ($1 \lesssim M / M_{\odot} \lesssim 100$), and it shows that black holes can carry significant angular momentum ($J \lesssim M^2$). In the 2030s and beyond, the space-based observatories such as the Laser Interferometer Space Antenna (LISA) \cite{LISAproposal, LISA:2024hlh}, TianQin/Taiji \cite{Gong:2021gvw}, deci-hertz detectors like DECIGO \cite{Kawamura:2020pcg}, will seek small-mass-ratio binaries, such as extreme- and intermediate-mass-ratio inspirals (EMRIs and IMRIs), among other sources, in which both the primary (e.g. a galactic black hole $M_1 \sim 10^{7} M_{\odot}$) and secondary (e.g., $M_2 \sim 10 - 10^4 M_{\odot}$) bodies are spinning, and for which the orbit remains eccentric and non-planar up until the capture of the smaller body \cite{Hu:2024vvu,LISA:2022yao}.

Successful parameter estimation requires accurate modelling of EMRIs and IMRIs across a wide parameter space through $\sim 10^3 - 10^6$ orbital cycles, to within one radian of phase~\cite{LISAConsortiumWaveformWorkingGroup:2023arg}. This is widely expected to demand waveforms at first post-adiabatic order, i.e., including the first subleading correction in the mass-ratio expansion of the waveform phase within gravitational self-force theory~\cite{Hinderer:2008dm,Poisson:2011nh,BaPo.18,PoWa.21}, which in turn requires carrying the self-force expansion on a Kerr background through second order in the mass ratio~\cite{vandeMeent:2017bcc,Nasipak:2025tby, Pound:2012nt,Wardell:2021fyy,Leather:2026gnc}. At the same order, the spin of the secondary enters the phase through a spin--curvature coupling~\cite{Mathews:2025nyb}, contributing both a conservative correction to the orbit and a dissipative correction to the
fluxes~\cite{Drummond:2022efc, Drummond:2022xej,Skoupy:2023lih,Drummond:2023wqc, Pio.25,Skoupy:2025nie}. Its contribution is comparable to that of the second-order self-force~\cite{Po.al.20,War.al.21}, and has spurred rapid progress toward waveform models for spinning secondaries on generic orbits
around spinning primaries~\cite{Drummond:2023wqc,Mathews:2025txc,Honet:2025lmk,Drummond:2026haw,Skoupy:2026ewu,Cui:2026qsk}. Separable equations of motion, reducible to quadratures, are particularly valuable in this context: they allow fast and accurate evaluation of the orbits that enter waveform models~\cite{Chua:2020stf,Katz:2021yft,Speri:2023jte,Chapman-Bird:2025xtd}, and provide analytical insight into the dynamics.

In this work, we focus on the conservative part of this problem at the test-body level, neglecting self-force effects, including radiation reaction. On Kerr spacetime, the post-geodesic effects of the spin-curvature coupling acting on the secondary have been investigated by several authors \cite{WitzHJ.19, Drummond:2022efc, Drummond:2022xej, Piovano:2025aro,Piovano:2026wpz}. Notably, Skoup\'y and Witzany~\cite{Skoupy:2024uan} have shown that, at linear order in spin, the physical worldline can be related to a nearby fiducial geodesic with shifted constants of motion, by making use of the hidden symmetry of Kerr spacetime~\cite{Yasui:2011pr,Frolov:2017kze}; the equations of motion then separate as in the geodesic case~(see also Ref.~\cite{Drummond:2026haw} for a related idea). 
The hidden symmetry underlying this construction is reviewed in Sec.~\ref{sec:intro-hidden}; it is shared by a wider class of spacetimes than Kerr.

\subsection{The motion of spinning body in curved spacetime}

A generic extended body in curved spacetime can be, when it is sufficiently compact, modelled as a test small body endowed with a finite set of multipole moments. The dynamics of the body is then described by an equation of motion for the body's representative world line, along with the evolution equation for the multipole moments.  
The Mathisson–Papapetrou-Tulczyjew–Dixon (MPTD) equations are the evolution equations for the body's multipole moments. At leading monopole order in the multipolar expansion, one recovers the geodesic approximation (neglecting the self-force effects): the smaller body traces a geodesic orbit in the spacetime associated with the larger object (say, a Kerr black hole). At subsequent orders, the expansion accounts for the body’s spin and higher order multipoles. The foundations of this multipolar description were laid by Mathisson \cite{Ma.37}, Papapetrou \cite{Pa.51}, and Tulczyjew \cite{Tu.59}, culminating in Dixon's framework \cite{Di.74}, which holds for any sufficiently small object with a conserved stress-energy-momentum tensor \cite{Di.74, Ha.15}, and Harte's modern formulation using generalized Killing fields \cite{Ha.12, Ha.15, Ha.20} which accounts for the self-field effects. In our work, however, self-field effects and back-reaction from the body onto the background spacetime are neglected.

\subsection{Hidden symmetry: special tensors in a curved spacetime}
\label{sec:intro-hidden}

Investigations of particle dynamics on black hole spacetimes have revealed the role played by Killing--St\"ackel (KS) tensor and Killing-Yano (KY) tensors~\cite{Yano.52} (cf.~reviews in Refs.~\cite{Yasui:2011pr,
Frolov:2017kze}). In 1968, Carter \cite{PhysRev.174.1559} showed that the Hamilton-Jacobi equation for geodesic motion on Kerr is integrable: there exists a fourth constant of motion which is quadratic in the momentum, $\GC_0 = K^{ab} p_a p_b$ (where $p_a = g_{ab} \dot{x}^b$). This implies that $K^{ab}$ is a KS tensor, i.e.~it satisfies $\nabla_{(a} K_{bc)} = 0$. Floyd \cite{Floyd:1973:dynamics} with Penrose \cite{Penrose:1973um} showed that the KS tensor can be written as the ``square'' of an (antisymmetric) KY tensor $f_{ab}$ whose covariant derivative is a three-form (i.e.~$\nabla_a f_{bc}$ is fully antisymmetric). 
While a Killing tensor suffices for geodesic separability, spinning bodies require the stronger KY structure: it follows immediately that the vector $f^{ab} p_b$ is parallel-transported along a geodesic, and thus the gyroscopic problem of finding a parallel-transported basis reduces to finding a single angle by quadrature \cite{kamran1986parallel}. 

A KY tensor can be viewed as more fundamental than a KS tensor \cite{Frolov:2006dqt} because its ``square'' is always a KS tensor, but not all KS tensors can be decomposed in this way \cite{Ferrando:2002dp}. In four-dimensional spacetimes, Collinson \cite{Collinson:1976:relationship} showed that, if a vacuum solution of the Einstein equations admits a nondegenerate KY tensor, then it is of Petrov type D. A complete classification of vacuum type D solutions in four dimensions was given by Kinnersley \cite{Kinnersley:1969zza}. In the absence of acceleration, it was shown by Demianski and Francaviglia \cite{Demianski:1980mgt} that these solutions all admit Killing and KY tensors; however, the accelerated cases do not allow KY tensors.

More recently, work on higher-dimensional black holes led to deeper understanding of the role of the \emph{principal CKY tensor} $f^\star_{ab}$: a closed conformal KY tensor, which in 4D is the Hodge dual of $f_{ab}$ \cite{Frolov:2017kze}. It was established in Ref.~\cite{Krtous:2008tb} that the metric of any higher-dimensional spacetime which admits a principal CKY tensor can be put into a canonical form, which becomes the Kerr-NUT-(A)dS metric once the Einstein equations with the cosmological constant are imposed. The canonical form guarantees several properties: (1) it is of the algebraic type D, (2) it allows a separation of variables for the Hamilton–Jacobi, Klein–Gordon, and Dirac equations, and (3) the geodesic motion in such a spacetime is completely integrable. Follow-up work showed that Maxwell's equations are also separable \cite{Krtous:2018bvk}, and the gyroscopic problem simplifies considerably \cite{Connell:2008vn}. Note that the classification results above are formulated for vacuum solutions, whereas the construction in this paper only requires the existence of a non-degenerate KY tensor and does not rely on the Einstein equations.

Beyond geodesic motion, at linear order in spin, constants of motion constructed from the KY tensor are known to exist \cite{Rudiger.I.81, Rudiger.II.83, ComDru.22,Ra.Iso.IntegO1.26,Andersson:2025bhq}, and the resulting system is integrable~(see also Refs.~\cite{Gibbons:1993ap,Kubiznak:2011ay} for a related model). They have been used to separate the Hamilton--Jacobi equation \cite{WitzHJ.19, Witzany:2026eqc} and to obtain analytic solutions in Schwarzschild and Kerr \cite{TheocharisAApostolatos_1996,WiPio.23,Ra.Iso.PapII.24,Skoupy:2024uan,Piovano:2025aro,Piovano:2026wpz}. Linear order in spin is the first step beyond Kerr geodesics at which, outside of Kerr, the solutions to the equations of motion are not known; in this paper we provide them for spacetimes admitting a non-degenerate KY tensor, beyond Kerr, beyond vacuum and beyond geometries satisfying the Einstein equation.
At quadratic order in spin, by contrast, integrability is sensitive to the internal structure of the body, and is known to hold only for black holes~\cite{Ra.CQG.24,Ra.Iso.Dru.IntegO2.26}; we restrict ourselves to linear order throughout.

The purpose of the current work is to generalise the worldline-shift result to this wider class of spacetimes, to establish separability, and to show that the motion of a spinning body maps onto that of a fictitious charged particle subject to an effective, curvature-induced Lorentz force. Throughout, we centre the role of the Killing--Yano tensor, which enables covariant derivations that do not rely on any specific coordinate system.

\subsection{Summary of main results}

The test body's motion is described by the MPTD equations \cite{Ma.15,Pa.51,Tu.57,Di.64} at dipolar order \cite{EhRu.77,ObuPue.11,GraHarWal.10}, truncated at linear order in spin. The equations describe the evolution of the body's linear momentum $p_a$ and spin tensor $S^{ab}$ along a representative worldline $\mathscr{W}$, via
\begin{equation*}
        \dot{p}_a    = R_{abcd} S^{bc} u^d,
			\quand
        \dot{S}^{ab} = 0\,,
\end{equation*}
where an overdot denotes the covariant derivative along the $\mathscr{W}$ and $u^a$ is the four-velocity. The worldline is fixed uniquely by the Tulczyjew--Dixon spin supplementary condition $p_aS^{ab} = 0$ \cite{Tu.59,Di.64}. If a Killing--Yano (KY) tensor $f_{ab}$ exists on the background spacetime, then there are two Killing vector fields $\xi^a$ and $\eta^a$ which generate the (normalised) invariants
\begin{equation*}
    E=-u_a\xi^a-\tfrac{1}{2}\nabla_a\xi_b s^{ab} \quand 
    {L}=u_a\eta^a+\tfrac{1}{2}\nabla_a\eta_bs^{ab}~,
\end{equation*}
with the normalised spin tensor $s^{ab}$ related to $S^{ab}$ in \eqref{eq:Killing-invariant}. These two quantities are exactly conserved along $\mathscr{W}$. In addition, there are three constants which are only conserved up to quadratic-in-spin corrections, namely
\begin{subequations}
    \begin{align*}
        \GC=K_{ab}u^au^b + 4u^as^{bc}f^d_{\phantom{d}[a}\nabla_d f_{c]b}~, \quad K = f_{ab}^\star s^{ab}~, \quand \mu^2 = -p_ap^a~. 
    \end{align*}
\end{subequations}
Here $\GC$ is a (generalised) Carter constant, $K$ measures the component of the body's spin along the orbital angular momentum, and $\mu$ is the body's mass. Together, these five invariants guarantee the integrability of the linear-in-spin system when written as a Hamiltonian system \cite{WitzHJ.19,Ra.Iso.IntegO1.26}. Importantly, the aforementioned results only rely on the existence of $f_{ab}$, not on the metric being Ricci-flat, or even a solution to Einstein's equations. 

In this work, we illustrate how the KY symmetry can be used to map the real, physical worldline $\mathscr{W}$ to a fictitious worldline, $\mathscr{V}$, with tangent $\nu^a$ satisfying
\begin{equation*}
    \nu^b\nabla_b\nu^a = q F^a_{\phantom{a}b}\nu^b~,
\end{equation*}
i.e., $\mathscr{V}$ is the worldline of a fictitious spinless particle of charge $q$, subject to a Lorentz force from an effective field strength $F_{ab}$.
Most importantly, all these fiducial objects $(\mathscr{V},q,\nu^a,F^{ab})$ are defined explicitly in terms of the real, physical background and original spinning particle, in particular the charge is $q\propto K/Q$, and the field strength tensor is $F_{ab} \propto f^\star_{ac}R^c_{\phantom{c}b}$. The latter vanishes exactly for Ricci-flat spacetimes, in which case $\mathscr{V}$ is simply a geodesic. 
Whether or not the spacetime is Ricci-flat, four constants of motion exist along $\mathscr{V}$, namely
\begin{subequations}
    \begin{align*}
        \mathscr{E}=-\left(\nu_a+qA_a\right)\xi^a , \qquad 
        \mathscr{L} =\left(\nu_a+ q A_a\right) \eta^a , \qquad 
        \widetilde \GC_0 =K_{ab}\nu^a\nu^b,
\end{align*}
\end{subequations}
and the tangent vector is normalized $\nu_a\nu^a=-1$. This guarantees the integrability of the resulting dynamics~\cite{Ra.Iso.IntegO1.26}. Together with the canonical coordinates of Ref.~\cite{Frolov:2017kze}, they render the equations of motion on $\mathscr{V}$ separable, cf.~\eqref{eq:nu-sep}, as is familiar for charged particles.

The construction involves three curves. The physical worldline $\mathscr{W}$ is first shifted, by a vector built from the Killing--Yano tensor and the Marck tetrad, to a nearby worldline $\tilde{\mathscr{W}}$. Its tangent $\tilde u^a$ differs from $\nu^a$ only by a constant rescaling and a drift along the primary Killing vector, so $\tilde{\mathscr{W}}$ and $\mathscr{V}$ trace the same path in the remaining coordinates. Working back from $\mathscr{V}$ to $\tilde{\mathscr{W}}$, and then undoing the shift, the physical worldline $\mathscr{W}$ likewise reduces to quadratures, cf.~\eqref{eqn:Gen-Sep-Sols}.

Our construction is completely off-shell, in the sense of \cite{Frolov:2017kze}: the background metric is only asked to admit a non-degenerate KY tensor, and not be a vacuum metric, nor a solution to Einstein's field equations. In the case of a Kerr background, our results recover those found in \cite{Skoupy:2024uan}.
We consider Kerr-de Sitter spacetime in Boyer-Lindquist coordinates $(t,r,z,\phi)$ with $z = \cos{\theta}$ as an example of a non-Ricci-flat geometry. It explicitly provide separable equations for the shifted worldline $\tilde{\mathscr{W}}$, parametrized by the deformed Carter--Mino time $\tilde\lambda$ of Ref.~\cite{Skoupy:2024uan}
(see Sec.~\ref{sec:Kerr-dS example}), viz.
\begin{subequations}
\begin{align}
      &\frac{\mathrm{d}\tilde t}{\mathrm{d}\tilde\lambda}
      = -\frac{X_z}{1+\Lambda a^2z^2/3}
         - \frac{(r^2+a^2)\,X_r}{\Delta}
         - \frac{3\tilde{\Sigma}\,s_\parallel }{2\sqrt{\GC_0}}, \quad \dv{\tilde{\phi}}{\tilde{\lambda}} 
      = -a \chi \left(\fr{X_z(z)}{\Theta(z)} + \fr{X_r(r)}{\Delta(r)}\right), \nonumber \\
      &a^2\left(\dv{\tilde{z}}{\tilde{\lambda}}\right)^2 =\Theta(z)\left(\tilde{\GC}_0 - a^2 z^2\right)  - X_z^2(z), \quad
      \left(\dv{\tilde{r}}{\tilde{\lambda}}\right)^2 = X_r^2(r) - \Delta(r) \left(\tilde{\GC}_0 + r^2\right). \nonumber
\end{align}
\end{subequations}
The integrability of the linearised system has been shown in complementary ways (see \cite{Ra.Iso.IntegO1.26, Ra.Iso.PapII.24, Ra.CQG.24} for discussions using Hamiltonian and symplectic methods, and \cite{Witzany_2025, Witzany:2019nml} for discussions using the Hamilton-Jacobi equation), all of which rely on an underlying KY symmetry. Our work provides a concrete realisation of this integrability to provide separable equations of motion independent of the Einstein field equations. In the context of EMRI and IMRI modelling, one can now imagine carrying out the waveform generation pipeline using these solutions as input to include secondary spin effects for spacetimes beyond Kerr, with matter, or arising in modified gravity. 

\subsection{Organization of the paper}

Our presentation is organised as follows. In Sec.~\ref{sec:spin-bodies}, we briefly review the linear-in-spin dynamics of spinning test bodies in spacetimes admitting a Killing--Yano tensor. In Sec.~\ref{sec:cov-shift}, we give a covariant construction of the worldline shift, generalising the Kerr proposal of Ref.~\cite{Skoupy:2024uan}. We show that, for a suitable choice of the shift, the perpendicular spin leaves the constants of motion unchanged, while the parallel spin leads naturally to an interpretation in terms of an effective charge. In Sec.~\ref{sec:eff-eom}, we make this interpretation precise. We construct the effective gauge field, and show that in the off-shell metric the motion separates and obeys a Lorentz-force equation. We then illustrate the formalism in Kerr--de Sitter spacetime (Sec.~\ref{sec:Kerr-dS example}). We conclude in Sec.~\ref{sec:discussion},
and collect conventions and detailed proofs in the appendices.

\section{Spinning test bodies and Killing-Yano symmetries}
\label{sec:spin-bodies}

This section collects the ingredients on which the construction rests. We first recall the linearised MPTD equations with the Tulczyjew--Dixon spin supplementary condition (Sec.~\ref{sec:MPTD}), which at linear order in spin reduce to $p^a=\mu u^a$ and a parallel-transported spin vector $s^a$. We then introduce various Killing objects and list identities satisfied by them (Sec.~\ref{sec:KY}) as well as the conserved quantities that they generate (Sec.~\ref{sec:conserved}). Lastly, we discuss the Marck tetrad and decompositions into parallel and perpendicular components (Sec.~\ref{sec:Marck}).

\subsection{Linear-in-spin MPTD dynamics}
\label{sec:MPTD}

In relativistic dynamics, a small extended body moving in a fixed spacetime with metric tensor $g_{ab}$ can be described covariantly, and in closed form, as a point particle carrying multipole moments. We denote the coordinates covering the four-dimensional spacetime $x^{\alpha}$. 
That such a ``reduced multipole description'' exists at all is far from obvious \cite{EhRu.77}. Its existence has nevertheless been established along several independent routes: multipolar expansions \cite{Tu.59,Di.74,StPu.10}, Lagrangian methods \cite{St.15}, self-force methods \cite{Mino:1997bw,Po.al.11,Po.12,Gr.12,MaPoWa.22}, and generalized Killing fields \cite{Ha.12,Ha.15,Ha.20,Harte:2025tmd}. Here we follow, and only sketch, the Dixon--Harte construction \cite{Di.74,Di.15,Ha.12,Ha.15}.

Given a spatially compact, conserved stress-energy tensor $T^{ab}$ and a reference worldline $\mathscr{W}$, one defines moments of $T^{ab}$ on a family of hypersurfaces foliating the body's worldtube. The two lowest ones, the monopole $I^{ab}=I^{(ab)}$ and the dipole $I^{abc}=I^{a(bc)}$, are conveniently rewritten as
\begin{equation} \label{IspSJ}
    I^{ab} = p^{(a} V^{b)} \quand I^{abc} = S^{a(b} V^{c)},
\end{equation}
where $V^a$ is tangent to $\mathscr{W}$. These relations define the linear momentum $p^a$ and the spin tensor $S^{ab}=S^{[ab]}$, so that the particle carries $4+6=10$ degrees of freedom at dipolar order. The conservation law $\nabla_aT^{ab}=0$ is then equivalent to the Mathisson--Papapetrou--Tulczyjew--Dixon (MPTD) equations \cite{Ma.15,Pa.51,Tu.57,Di.64}
\begin{subequations} \label{EEgen}
\begin{align}
	\dot{p}_a  &= R_{abcd} S^{bc}u^d + F_a, \label{EoM} \\
	\dot{S}^{ab} &= 2 p^{[a} u^{b]} + N^{ab}, \label{EoP}
\end{align}
\end{subequations}
where $\dot{X}:=u^a\nabla_aX$ denotes the covariant derivative along the worldline, $u^a$ is the four-velocity along $\mathscr{W}$ and $R_{abcd}$ is the curvature of the background. 
The force $F_a$ and torque $N^{ab}$ collect the couplings of the quadrupole and higher multipoles to curvature \cite{Di.74,Ha.12}. The first terms on the right-hand sides are pure {kinematics}, already present in Newtonian mechanics \cite{Ha.23}: the spin--curvature coupling drives $p^a$, and the misalignment of $p^a$ and $u^a$ drives $S^{ab}$. Quadrupolar and higher-order effects are, by contrast, genuinely {dynamical}. As in Newtonian mechanics, no evolution equation for these higher moments follows from the formalism: their dynamics depends on the internal structure of the body and must be supplied by a constitutive model.

In this work we truncate the multipole expansion at dipolar order and work \emph{linearly in the spin}. For a spinning body, the quadrupole induced by rotation scales as $\text{spin}^2$, so that $F_a$ and $N^{ab}$ are of second order in spin, as are the non-linearities that the dipole alone would generate. Retaining the latter while discarding the former would not be order-consistent. At linear order, the only effect of the spin is therefore the spin--curvature coupling.

Three scalar fields can be built along $\mathscr{W}$ from $(p_a,S^{ab})$:
\begin{equation}\label{eq:def-norms}
    \mu^2 := -p_a p^a, \quad
    S_\circ^2 := \frac{1}{2}S_{ab}S^{ab} \quand
    S_\star^2 := \frac{1}{8}\varepsilon_{abcd}S^{ab}S^{cd},
\end{equation}
where $\varepsilon_{abcd}$ is the Levi-Civita tensor. They are the dynamical mass, the spin norm, and a pseudo-scalar measuring the mass-dipole content of $S^{ab}$, respectively. They are intrinsic to the particle and depend on no observer, but they are not conserved in general: their evolution depends on the choice of representative worldline and on $(F_a,N^{ab})$.

Even with $F^a=0$ and $N^{ab}=0$, the differential system \eqref{EEgen} is underdetermined: $(u^a,p_a,S^{ab})$ comprise thirteen unknowns (since $u^au_a=-1$) for only ten equations. 
A \emph{spin supplementary condition} (SSC) adds three algebraic constraints and fixes the representative worldline \cite{Be.67,CoNa.15}, or equivalently the origin of the multipole moments. We adopt the Tulczyjew--Dixon (TD) SSC \cite{Tu.59,Di.64}
\begin{equation}\label{TDSSC}
    p_a S^{ab} = 0 \,.
\end{equation}

Together with the MPTD equations, the TD SSC yields an exact relation between $p^a$ and $u^a$. At dipolar order it reads \cite{EhRu.77,ObuPue.11,GraHarWal.10}
\begin{equation}\label{momvelTD}
    \mu u^a = p^a + \frac{1}{2\mu^2} S^{ab} R_{bcde} p^c S^{de}.
\end{equation}
The correction is quadratic in spin: it is the so-called \emph{hidden momentum}, a purely relativistic effect. At linear order, it reduces to the familiar proportionality 
\begin{equation}\label{momvellin}
     p^a = \mu u^a + O(S^2), 
\end{equation}
which therefore holds, just as in Newtonian gravity \cite{Ha.15,HaGa.20}. We note that the same TD-SSC, MPTD equations give the evolution equation for the scalar fields 
\begin{equation}
    \frac{\ud \mu}{\ud \pt} = O(S^2) \quand \frac{\ud S_\circ}{\ud \pt} = O(S^2)\,.
\end{equation}
Therefore $\mu$ and $S_\circ$ are conserved along ${\mathscr{W}}$ at this order \cite{Gr.al2.10}, 
while we have $S_\star=0$ identically~\cite{WiStLu.19,Ra.Iso.IntegO1.26}. 

Under the TD SSC, we can also construct the equivalent spin vector $S^a$ from the spin tensor $S^{ab}$ as 
\begin{equation}\label{eq:def-TD-spin}
\quad S^b := \frac{1}{2} \varepsilon^{abcd}\, u_a S_{cd} \quad 
{\Longleftrightarrow} \quad S^{ab} = \varepsilon^{ab}_{\phantom{ab}cd} u^c S^d, 
\end{equation}
such that $S^a$ is the spin 4-vector measured by an observer of four-velocity $u_a$. The TD spin vector is found to satisfy
\begin{equation}\label{eq:cons-Casimir}
    {p}_{a} S^{a} = 0 
    \quand
    S^{a}\,S_{a} = S_{\circ}^2 \,.
\end{equation}

Collecting the above, we shall work from now on with the following algebraic-differential system: linearized MPTD equations with the TD SSC in Eq.~\eqref{TDSSC} 
\begin{equation}\label{EElin}
        \dot{p}_a    = R_{abcd} S^{bc} u^d, 
		\quand
        \dot{S}^{ab} = 0\,. 
\end{equation}
and, as a corollary, we have (cf. Ref.~\cite{GraHarWal.10}) the evolution equation for the spin vector
\begin{equation}\label{eq:dot_TD_Spin}
    \dot{S}^a = 0\,.
\end{equation} 
Two consequences are worth stressing. First, $S^{ab}$ and $S^{a}$ are parallel-transported along $\mathscr{W}$: 
at the linear-order in spin, it is parallel-transported along the background geodesic. 
Second, the only departure from geodesic motion is a spin-curvature acceleration, $\mu\,\dot u_a = R_{abcd} S^{bc} u^d$. 

In what follows, any term involving two or more powers of spin $S^{ab}$ is discarded.

\subsection{Hidden symmetries: Killing--Yano tensor}
\label{sec:KY}

We assume that the spacetime admits a non-degenerate KY two-form $f_{ab}$. It is an antisymmetric tensor field (2-form) satisfying 
\begin{equation}
    \nabla_{(a} f_{b)c}=0 \quand \mathfrak{Z}:=\frac{1}{4} f^\star_{ab}f^{ab}
    \neq0,
    \label{eq:KY-def}
\end{equation}
where the first equation is the defining KY equation, and the second is the non-degeneracy condition. In four dimensions, this condition means that $f_{ab}$ has the (matrix) rank 4, in the sense of R\"{u}diger \cite{Rudiger.I.81,Rudiger.II.83}.

We define the Hodge dual of the KY tensor by
\begin{equation}
f_{ab}^\star:=\frac{1}{2}\,\varepsilon_{ab}{}^{cd}f_{cd},
    \label{eq:KY-dual}
\end{equation}
where $\varepsilon_{abcd}$ is the totally antisymmetric Levi-Civita tensor. In Lorentzian spacetimes it follows that $f^{\star\star}_{ab}=-f_{ab}$. Both tensors combine into the algebraic identity 
\begin{equation}
    f^\star_{ac} f^c{}_b = -\mathfrak{Z}\, g_{ab}.
    \label{eq:KY-algebraic}
\end{equation}

In a four-dimensional spacetime, the dual $f^\star_{ab}$ is a non-degenerate Closed Conformal Killing--Yano tensor, also known as the principal tensor (see, e.g., reviews \cite{Yasui:2011pr,Frolov.17}). The KY dual satisfies a differential relation akin to \eqref{eq:KY-def} for $f_{ab}$, namely
\begin{equation}
    \nabla_a f^\star_{bc}
    =
    2g_{a[b}\,\xi_{c]},
    \label{eq:KY-principal-derivative}
\end{equation}
where the vector $\xi^a$ is a primary Killing vector uniquely associated with $f_{ab}$ via 
\begin{equation}
    \xi_a=\frac13\nabla^b f^\star_{ba}
    \quad \Longleftrightarrow \quad
    \nabla_a f_{bc}=-\varepsilon_{abcd}\xi^d 
    .
    \label{eq:KY-xi}
\end{equation}
This Killing vector is also involved in the derivative of the scalar $\mathfrak{Z}$: taking a covariant derivative of \eqref{eq:KY-algebraic} and using metric compatibility gives the simple relation
\begin{equation}
    \nabla_a \mathfrak{Z} = f_{ab}\,\xi^b.
    \label{eq:KY-gradZ}
\end{equation}
The KY tensor also generates a symmetric Killing--St\"{a}ckel (KS) tensor,
\footnote{Some works define the Killing-St\"{a}ckel tensor by contracting on the other index, i.e., $K_{ab}=f_{ac} f^{c}_{\phantom{c}b}$. This is purely conventional.}
%
\begin{equation}
    K_{ab} := f_{ac} f_b{}^c,
    \quad \Rightarrow \quad
    \nabla_{(a}K_{bc)}=0.
    \label{eq:KY-K}
\end{equation}

From the KS tensor, a second Killing vector is generated by contracting it with $\xi^a$:
\begin{equation}
    \eta^a := K^{ab}\,\xi_b,
    \quand
    \nabla_{(a}\,\eta_{b)}=0.
    \label{eq:KY-eta}
\end{equation}

In Kerr spacetime, $\xi^a$ is proportional to the stationary Killing vector, while the axial Killing vector is a linear combination of $\xi^a$ and $\eta^a$. The normalisations of the physical energy and angular momentum therefore depend on the chosen linear combination.

Several further identities follow from the integrability conditions of the KY equation in Eq.~\eqref{eq:KY-def}. The curvature expression for the derivative of $\xi_a$ is
\begin{equation}
    \nabla_b\xi_a
    =-\frac14 R_{ab}{}^{cd}f^\star_{cd}
      +\frac12 R_a{}^c f^\star_{cb}.
    \label{eq:KY-dxi}
\end{equation}
In particular, the existence of a KY tensor imposes the curvature constraints
\begin{equation}
    R_{(a}{}^c f_{b)c}=0,
    \quand
    R_{(a}{}^c f^\star_{b)c}=0.
    \label{eq:KY-Ricci-constraints}
\end{equation}
The derivative of the secondary Killing vector can be expressed as 
\begin{equation}
    \nabla_a\eta_b=
       \varepsilon_{abde}\xi^e f^{dc}\xi_c
       +\frac14 f_a{}^f f_b{}^c R_{fcde}f^{\star de} +\frac{\mathfrak Z}{2}R_a{}^c f_{bc}.
\label{eq:KY-deta}
\end{equation}

The existence of a non-degenerate KY tensor implies that the background's Weyl tensor is of Petrov type D, and imposes other constraints on the Ricci tensor \cite{DiRu.I.81,DiRu.II.82,Ra.Iso.Dru.IntegO2.26}. For the calculations below, it is useful to express the Killing vector derivatives \eqref{eq:KY-dxi}-\eqref{eq:KY-deta} in terms of two curvature scalars: the Weyl scalar $\Psi_2$ and the Ricci scalar $R$. These expressions are (see Appendix~\ref{app:ids} for derivation)
\begin{subequations}
    \begin{align}
    \label{eq:KY-dxi-typeD}
    \nabla_b\xi_a
    &=\left(\mathrm{Re}[\Psi_2]+\frac{R}{12}\right)
       f^\star_{ab}
      -\mathrm{Im}[\Psi_2]\,f_{ab}, \\ 
    \label{eq:KY-deta-typeD}
    \nabla_{b} \eta_{a} &= \mf{Z}\left(\mathrm{Re}[\Psi_2] - \frac{R}{12}\right)f_{ab} - \mathrm{Im}[\Psi_2]K_a^{\ph c} f_{cb} - 2\xi^c f^\star_{c[a}\xi_{b]} - \xi^2 f^\star_{ab} \,.
    \end{align}
\end{subequations}
The trace-free part of the Ricci tensor does not appear in \eqref{eq:KY-dxi-typeD} and \eqref{eq:KY-deta-typeD}: it drops out by virtue of the curvature constraints~\eqref{eq:KY-Ricci-constraints}. 
The KY tensor thus generates the Killing tensor, two commuting Killing vectors \cite{frolov2011introduction,Frolov.17}, and the scalar and derivative identities needed to construct the conserved quantities. 

\subsection{Conserved quantities}
\label{sec:conserved}

\paragraph{Invariants associated with Killing vectors.}
For any Killing vector $k^a$, the MPTD equations leave invariant the following scalar field
\begin{equation}
    \mathcal F_k := u_a k^a + \frac{1}{2} s^{ab}\nabla_a k_b\,
    \quad \text{with} \quad
    s^{ab} := \frac{S^{ab}}{\mu}\,,
    \label{eq:Killing-invariant}
\end{equation}
where $s^{ab}$ denotes the spin tensor in the normalisation used for the tangent-based equations. It generalises the geodesic invariant $u_a k^a$ by including the coupling between the spin and the derivative of the symmetry generator. This exact conservation law, which holds for arbitrary multipolar bodies, is not a consequence of the pole--dipole MPTD equations alone, but can be shown in the broader Dixon-Harte framework \cite{Di.74,Ha.12,Ha.15,Di.15}.

Applying this construction to the KY-generated Killing vectors $\xi^a$ \eqref{eq:KY-xi} and $\eta^a$ \eqref{eq:KY-eta}, we obtain two invariants defined in a normalized form as
\begin{subequations}\label{eqn: KY-MPTD-PCQ}
\begin{align}
    E &= -\xi_a u^a-\frac{1}{2} \nabla_a \xi_b \,s^{ab},
    \label{eq:KY-energy}\\
    L &= \eta_a u^a+\frac12\nabla_a\eta_b\,s^{ab}.
    \label{eq:KY-angular-momentum}
\end{align}
\end{subequations}
The signs and normalisations in these expressions follow the conventions of the present work. The quantity $L$ here is a natural invariant coming from the second Killing vector. It need not be the usual axial angular momentum in a generic KY spacetime. In Kerr, that axial angular momentum is an appropriate linear combination of $E$ and $L$. 

\paragraph{The (generalized) Carter constant.}
The KS tensor generates a quadratic invariant for geodesic motion, known as the Carter constant. For a spinning body, the corresponding quantity acquires a term linear in the spin. The (generalized) Carter constant is defined by 
\begin{equation}
    \GC := \GC_0 - 2u^a s^{bc}L_{abc}\,,
    \label{eq:KY-Carter}
\end{equation}
with    
\begin{equation}\label{eq:def-GC0-L}
    \GC_0 := K_{ab}u^a u^b 
    \quand 
    L_{abc}:=-2f^d{}_{[a}\nabla_d f_{c]b}\,.
\end{equation}
The first term $\GC_0$ is the (well-known) geodesic part of the Carter constant~\cite{Carter.68}. The second term involving $L_{abc}$ is the spin correction required for conservation under the linear-in-spin MPTD equations. With the equations of motion and the TD SSC imposed, this quantity obeys
\begin{equation}
    \frac{\ud\GC}{\ud\pt} = O(s^2).
    \label{eq:KY-Carter-conservation}
\end{equation}

Thus $\GC$ is a quasi-invariant \cite{ComDru.22,Ra.Iso.IntegO1.26}: it is conserved to the order retained here, rather than being an exact invariant of the full finite-spin dynamics. \\

\paragraph{The R\"{u}diger invariant and parallel spin.} A second KY-related quasi-invariant is the R\"{u}diger invariant, which can be written as
\begin{equation}
    K := f^\star_{ab}s^{ab}.
    \label{eq:KY-Rudiger}
\end{equation}
It is conserved through linear order in spin,
\begin{equation}
    \frac{\ud K}{\ud \pt} = O(s^2).
    \label{eq:KY-Rudiger-conservation}
\end{equation}

Under the TD SSC, the R\"{u}diger invariant admits an interpretation as the component of the spin along the orbital angular momentum. The KY tensor defines a covariant angular-momentum four-vector~\cite{DiRu.I.81,DiRu.II.82} (we adopt the sign convention of~\cite{DruHug.I.22}),
\begin{equation} \label{eq:def-SDL}
\mathcal{L}^b := p_a f^{ab}.
\end{equation}
This vector ${\mathcal {L}^a}$ has the dimensions of (orbital) angular momentum, and the squared norm of $\mathcal{L}^a$ reduces to the Carter constant in the geodesic limit, $\mathcal{L}^a \mathcal{L}_a = K_{ab} p^a p^b + O(s)$. In fact, it reduces to the familiar $\mathbf{r}\times\mathbf{p}$ (up to the overall normalization) when evaluated in a natural basis in Schwarzschild spacetime~\cite{DruHug.I.22,Ra.Iso.PapII.24}.

Inserting the inverse relation~\eqref{eq:def-TD-spin} between the spin vector and the spin tensor into~\eqref{eq:KY-Rudiger}, we find 
\begin{equation}\label{eq:Rudiger-SO}
\mathcal{L}^a s_a = \mu K.
\end{equation}
The R\"{u}diger invariant thus measures the projection of the spin onto $\mathcal{L}^a$, which justifies the ``parallel-spin'' description.

The body's mass, energy, second Killing-vector invariant, Carter constant,  and R\"udiger (or parallel-spin) invariant supply the conserved quantities used in the linear-in-spin integrability construction of the MPTD dynamics under the TD SSC in a KY spacetime~\cite{Ra.Iso.IntegO1.26}. 

All statements concerning $\GC$ and $K$ are understood modulo terms quadratic in spin.

\subsection{The Marck tetrad and a linear in spin solution}
\label{sec:Marck}

As shown in Eq.~\eqref{eq:dot_TD_Spin}, the spin vector $s^a$ is parallel transported along the MPTD worldline at linear order in spin. At this order, this is equivalent to parallel transport along the \emph{(background) geodesic} curve obtained by switching off the spin, whose tangent we denote by $v^a$. Along it, we use the approximated geodesic Carter constant $\GC_0 = K_{ab}v^a v^b$~\eqref{eq:def-GC0-L}; replacing $v^a$ by $u^a$ changes $\GC_0$ only at $O(s)$.

To solve this parallel transport, we introduce the orthonormal Marck tetrad $\left\{v, e_A \right\}$ for $A = 1,2,3$ \cite{Marck.83}, built directly from the KY tensor $f_{ab}$ and its Hodge dual, the principal tensor $f^\star_{ab}$. Its spatial legs are  
\begin{subequations}\label{eq:def-Marck-tetrad}
    \begin{align}
        (e_3)^a &= - \frac{1}{\sqrt{\GC_0}} f^{ab} v_{b}~, \label{eqn: Marck-tetrad-e3} \\ 
        (e_2)^a &= \frac{1}{\sqrt{\mathcal{N}}}\left(K^{ab} + \GC_0\,g^{ab} \right) v_b~, \label{eqn: Marck-tetrad-e2}\\
        (e_1)^a &= - \frac{1}{\sqrt{\GC_0}} f_\star^{ab} (e_2)_b~, \label{eqn: Marck-tetrad-e1}
    \end{align}
\end{subequations}
where $\mathcal{N}$ is a normalisation factor $\mathcal{N} = K_{ac}K^{c}{}_{b}\,v^a v^b + \GC_0^2$ ensuring $(e_2)^a (e_2)_a=1$.
The Marck tetrad has a clear geometric origin~\cite{KamMar.86}. The principal tensor $f^\star_{ab}$ singles out two orthogonal 2-planes, one timelike and one spacelike, as its eigenplanes. In Kerr spacetime in the Boyer--Lindquist coordinates $(t, r, \theta, \phi)$, these are spanned by the two principal null directions and by $\{\partial_\theta,\ a\sin^2\theta\,\partial_t+\partial_\phi\}$, respectively, and the eigenvalues of $f^{ab}_{\star}$ are given by the familiar coordinates $r$ and $a\cos\theta$.
Adapting an orthonormal frame to these planes yields the Darboux frame (on Kerr spacetime this is known as Carter's tetrad), in which $f_{ab}$ and $f^\star_{ab}$ are block-diagonal. Crucially, this construction makes no use of the field equations.
The Marck tetrad is obtained from it by a velocity-dependent Lorentz transformation: within each plane, the legs are boosted in the timelike plane and rotated in the spacelike plane to align with the projection of $v^a$, and the two planes are then mixed with coefficients depending only on the eigenvalues of $f^\star_{ab}$ and $\GC_0$. The only remaining freedom is a rotation in the $(e_1,e_2)$ plane.

Since $v_a s^a = 0$ from~\eqref{eq:cons-Casimir}, we may expand the spin vector by $s^a = s^A (e_A)^a$. In particular, the leg $(e_3)^a$ is parallel transported along the geodesic, as a consequence of the KY equation~\eqref{eq:KY-def} and the geodesic equation, so its component
\begin{equation}\label{eq:KY-s-parallel}
    s_\parallel := s_a (e_3)^a = -\frac{1}{\sqrt{\GC_0}}\, f^{ab} v_b s_a
\end{equation}
is a constant, which we call the parallel spin component; up to the normalisation, it is proportional to R\"{u}diger invariant $K$ in~\eqref{eq:Rudiger-SO}. 
Together with the conservation of the spin norm $s_\circ$~\eqref{eq:cons-Casimir},
this implies that the perpendicular spin component $
    s_\perp := \sqrt{s_\circ^2 - s_\parallel^2}$ is also constant.

One can combine $e_1^a$ and $e_2^a$ into a complex null vector 
\begin{equation}
    e_\perp^a = e^{-\ui\psi} \left( e_1^a + \ui\, e_2^a \right)\,,
\end{equation}
where $\psi=\psi(\pt)$ is the Marck phase. If $\psi$ evolves according to 
\begin{equation} \label{eq:Marck-angle-ODE}
    \frac{\ud \psi}{\ud \pt} = (e_{1})^a v^b \nabla_b (e_2)_a \,,
\end{equation}
then $e_\perp^a$ is parallel transported along the orbit. The expansion of $s^a$ now takes the form 
\begin{equation}\label{eqn: spin-vect-Marck}
    s^a = s_{||} e_3^a + s_\perp \rl{e_\perp^a}~, 
\end{equation}
which is manifestly parallel-transported as required, as long as the Marck phase $\psi$ satisfies the differential equation \eqref{eq:Marck-angle-ODE}. Equivalently, $s^1 = s_\perp\cos\psi$ and $s^2 = s_\perp\sin\psi$, with any initial phase absorbed into $\psi$.

Physically, the decomposition~\eqref{eqn: spin-vect-Marck} is interpreted as the parallel and perpendicular decomposition of the spin vector $s^a$. For later convenience, we also promote this parallel and perpendicular split of the spin vector, relative to the Marck tetrad to the spin tensor $s^{ab}$ using ~\eqref{eq:def-TD-spin}
(the subscript zero is again in anticipation of what follows). 
We write
\begin{equation}
    s^{ab} = s_{||} \mathfrak{F}_0^{ab} + s_\perp \rl{\mathfrak{G}_0^{ab}}
\end{equation}
with 
\begin{equation}\label{eq:def-frak-FG-geo}
    \mathfrak{F}_0^{ab} = \varepsilon^{ab}_{\phantom{ab}cd}v^c e_3^d 
    \quand 
    \mathfrak{G}_0^{ab} = \varepsilon^{ab}_{\phantom{ab}cd}v^c e_\perp^d~. 
\end{equation}
%

\section{The covariant worldline shift}
\label{sec:cov-shift}

In \cite{Skoupy:2024uan}, the authors propose a method to obtain analytic solutions for the (linearised) MPTD equations in Kerr spacetime by considering the MPTD worldline $x(\pt)$ as being ``near'' a shifted world-line $\tilde{x}(\pt)$. In this section, we construct this worldline-shift scheme in a covariant and coordinate-independent manner. 
Our construction applies to any spacetime admitting a non-degenerate KY tensor, and reduces to the Kerr results of Ref.~\cite{Skoupy:2024uan} as a special case.

Given the MPTD worldline $x(\pt)$ we consider a nearby worldline $\tilde{x}$, displaced from it by a shift vector $\shift^a$, i.e., $\tilde{x}^a = x^a + \shift^a$. Covariantly, the shift $\shift^a$ is the gradient at $x$ of Synge's world function between $x$ and $\tilde x$.
(For a brief introduction to the theory of bitensors, including the deviation $\sigma^a$ and the parallel-propagator $g^a~_\tsf{a}$ see Appendix \ref{app:shift-setup}). Note that for each point on $x(\pt)$ one can define a shift vector $\shift$ to many different worldlines. 
We seek a geometric prescription for the shift $\shift$, built the KY tensor $f_{ab}$, its dual $f^\star_{ab}$, and the Marck tetrad, such that the equations of motion for the resulting worldline separate.
This will be the goal of our approach. If $A_{ab}$ is a tensor on $x(\pt)$, after transporting to $\tilde{x}$, we will denote it as $\tilde{A}_{\textsf{ab}}$ with \textsf{sans serif} indices for objects on the new worldline. Then, $\tilde{A}_\tsf{ab}$ will depend on the parallel-propagator $g_\tsf{a}~^a$ and the deviation vector $\sigma_\tsf{a}$ and derivatives. 

For example, the worldline tangents are related as
\begin{equation}\label{eqn: two-tangents}
    u^a = {g^a}_\textsf{a} \left(\tilde{u}^\textsf{a} - \dot{\shift}^\textsf{a}\right)~,
\end{equation}
where we take ${g^a}_\tsf{a} = {\delta^a}_\tsf{a} + O(s^2)$. Importantly, we pick up a correction term from the shift for the spin tensor when evaluated on the new worldline \cite{Vines:2014oba,Vines:2016unv} as
\begin{equation}
    s^{ab} = {g^a}_\textsf{a}~{g^b}_\textsf{b}\left({\tilde s}^{\tsf{ab}} - 2 u^{[\tsf{a}}\shift^{\tsf{b}]}\right)~.
\end{equation}
To determine the shift such that the new equations of motion are separable (in some coordinate system), we expand the shift $\shift^a$ in a to-be-determined basis $\shift_A$ in exact analogy to the spin vector in the Marck basis \eqref{eqn: spin-vect-Marck}
\begin{equation}
    \shift^a = s_{||}\,\shift_3^a + s_\perp~\rl{\shift_\perp^a}~, 
\end{equation}
where $\shift_\perp^a = e^{-i\psi}\left(\shift_1^a + i \shift_2^a\right)$.
As the spin dependence is factored out front in terms of $s_{||}$ or $s_\perp$, it is sufficient to define the legs with respect to the geodesic trajectory $v^a$ at $O(s)$. In what follows,  whenever the spin dependence is factored out via $s_{||}$ or $s_\perp$ we will drop the $\tsf{sans serif}$ indices since we work at $O(s)$. We will now see how we can covariantly determine the shift vector(s).  

\subsection{Determining the shift vector}

Reconstructing the spin tensor along the new worldline yields a natural parallel/perpendicular split inherited from $s^a$,
\begin{equation}\label{eq:Marck-decomp-spin-tensor}
    \tilde{s}^{\textsf{ab}} = s_{||}\,\mf{F}^{\tsf{ab}} + s_\perp\rl{\mf{G}^{\tsf{ab}}}~,
\end{equation}
where
\begin{subequations}
\begin{align}
    \mf{F}^{ab} &= 2\,v^{[a}\shift_3^{b]} + \varepsilon^{ab}{}_{cd}\,v^{c}e_{3}^{d}~, \label{eqn: spin-tensor-split-par}\\
    \mf{G}^{ab} &= 2\,v^{[a}\shift_\perp^{b]} + \varepsilon^{ab}{}_{cd}\,v^{c}e_{\perp}^{d}~. \label{eqn: spin-tensor-split-perp}
\end{align}
\end{subequations}
Each sector consists of a mass-dipole part, generated by the shift, and a spin part. Relative to $v^a$, this has the same structure as the split of a Faraday tensor into electric and magnetic fields; in particular, the mass-dipole ``electric'' parts are recovered by projection, 
\begin{equation}\label{eq:shift-ele}
    \shift_3^a = \mf{F}^{ab}\,v_b \quand\shift_\perp^a = \mf{G}^{ab}\,v_b\,.
\end{equation}
Before the shift, the two sectors are orthogonal by construction (cf. Eq.~\eqref{eq:def-frak-FG-geo}), $\mathfrak{F}_{0\,ab}\mathfrak{G}_0^{ab}=0$, since the Marck tetrad is orthonormal. The shift adds mass-dipole parts to both, and we require that it preserve this orthogonality, so that the parallel and perpendicular sectors remain geometrically distinct. 
Since $\shift_3^a$ and $\shift_\perp^a$ are orthogonal to $v^a$, the only surviving contribution is $\mf{F}_{ab}\mf{G}^{ab} \propto (\shift_{3})_a\shift_\perp^a$, so the requirement reads 
\begin{equation}
    (\shift_3)_a \shift_\perp^a = 0. 
\end{equation}
This orthogonality is a first constraint on $\shift_3$ and $\shift_\perp$. Furthermore, the ``magnetic'' part $e_3$ is already known by \eqref{eqn: Marck-tetrad-e3}  as one of the spatial legs of the Marck tetrad,   $(e_3)^a = -f^{ab}v_b/\sqrt{\GC_0}$.  Using the identity
$\varepsilon_{dabc}f^{de} = \delta^e_a f^\star_{bc} - \delta^e_b f^\star_{ac} + \delta^e_c f^\star_{ab}$, Eq.~\eqref{eqn: spin-tensor-split-par} becomes
\begin{equation}
    X^{ab} = 2v^{[a}X^{b]c}v_c,
    \quad \text{with} \quad
    X^{ab} := \mathfrak{F}^{ab} + \frac{1}{\sqrt{\GC_0}}f^{\star ab}.
\end{equation}
This equation, however, states only that $X^{ab}$ has no ``magnetic'' part relative to $v^a$. We fix the remaining freedom by requiring that $\mathfrak{F}^{ab}$ carry no dependence on $v^a$ beyond the constant $\GC_0$; i.e., that it be built solely from geometric quantities inherent
to the spacetime and its Killing--Yano symmetry. This selects $X^{ab}=0$, namely, 
\begin{equation} \label{frakFequalsfstar}
    \mathfrak{F}^{ab} = -\frac{1}{\sqrt{\GC_0}}f^{\star ab}.
\end{equation}
Plugging this back to ~\eqref{eq:shift-ele}, $\shift_3$ is now determined to be
\begin{equation}
    \shift_3^a = -\frac{1}{\sqrt{\GC_0}}f_\star^{ab}v_b .
    \label{eqn: deriving-dx3}
\end{equation}
Since $\mathfrak{F}_{ab}\mathfrak{G}^{ab}=0$ and $\mathfrak{F}\propto f^\star$, the perpendicular sector must satisfy $\mathfrak{G}^{ab}f^\star_{ab}=0$. Evaluating this condition
(Appendix~\ref{app:deriv-dxperp}) fixes
\begin{equation}\label{eqn: deriving-dxperp}
    \shift_\perp^a = -\frac{\ui}{\sqrt{\GC_0}}\,f^a{}_b\,e_\perp^b .
\end{equation}
The shift is thus fixed entirely by the KY geometry, in any spacetime admitting a non-degenerate KY tensor. In Kerr, it can be shown that the expressions for the set $\left\{\shift_A\right\}$ and the parallel-spin tensor $\tilde{s}_\parallel^\tsf{ab}$ reduce to the expressions in \cite{Skoupy:2024uan}. 

\subsection{An auxiliary worldline and shifted constants of motion}
\label{sec: The-New-Worldline}
Taking a derivative of \eqref{eqn: two-tangents}, note that $\mathrm{D}u^a/d\pt$ satisfies the MPTD equations by definition. Then  $\mathrm{D}\tilde{u}^\textsf{a}/d\pt$ will also satisfy these equations only if $\dot{\shift}^\tsf{a}$ is parallel transported i.e. $\ddot{\shift}^\textsf{a} = 0$. This is the condition for the new worldline $\tilde{u}$ to be a physical MPTD trajectory. 
However, as we show in Appendix \ref{app:derivatives} (see also Appendix \ref{app:cov-proof}), $\ddot{\shift}_3^a$ and $\ddot{\shift}_\perp^a$ are orthogonal and no non-trivial linear combination would be zero. Hence $\ddot{\shift}^\tsf{a} \neq 0$. Then, the worldline $\tilde{x}$ cannot be a physical MPTD trajectory. Nevertheless, if one can obtain functional expressions for the components of $\tilde{u}^a$ which separate, one can always map this back to the original physical velocity $u^a$ via the shift. To make sense of this new trajectory, let us consider the original MPTD conserved quantities and the normalisation of the MPTD velocity, rewritten in terms of objects defined on $\tilde{x}$. Following \cite{Skoupy:2024uan}, we find that 
\begin{subequations}\label{eqn: tilde-conserved-quantities}
    \begin{align}
        E &= -\tilde{\xi}_\tsf{a}\left(\tilde{u}^\tsf{a} - \dot{\shift}^\tsf{a}\right) + \tfrac{1}{2}\tilde{\xi}_{\tsf{a;b}}\tilde{s}^{\tsf{ab}}~, \\
        {L} &= \tilde{\eta}_\tsf{a}\left(\tilde{u}^\tsf{a} - \dot{\shift}^\tsf{a}\right) - \tfrac{1}{2}\tilde{\eta}_{\tsf{a;b}}\tilde{s}^{\tsf{ab}}~, \\
        \GC &= \tilde{K}_{\tsf{ab}}\tilde{u}^\tsf{a}\left(\tilde{u}^\tsf{b}-2 \dot{\shift}^\tsf{b}\right) - 2\tilde{u}^\tsf{a}\tilde{s}^{\tsf{bc}}\tilde{L}_{\tsf{abc}}~, \phantom{\frac12} \\
        -1 &= \tilde{g}_{\tsf{ab}}\tilde{u}^\tsf{a}\left(\tilde{u}^\tsf{b} - 2 \dot{\shift}^\tsf{b}\right)~.
    \end{align}
\end{subequations}
As the new velocity is not a physical worldline, we seek to  rewrite the velocity as \emph{some} auxiliary vector $\nu$ plus additional terms, 
\begin{equation}\label{eq:nu-ansatz}
\tilde{u}^\tsf{a} = \nu^\tsf{a} + s_{||}~\alpha~ \dot{\shift}_3^\tsf{a} 
+ s_{\perp}~\beta~ \rl{\dot{\shift}_\perp^\tsf{a}}~, 
\end{equation}
where we have two unknown real coefficients $\alpha$ and $\beta$. 
Substituting this ansatz into \eqref{eqn: tilde-conserved-quantities} and regarding $\nu^a$ as tangent to a timelike curve, we define quantities with respect to $\nu^a$ analogous to those of the MPTD worldline. For instance, $\tilde E := -\tilde\xi_{\mathsf a}\nu^{\mathsf a}$ is related to the MPTD energy by
\begin{equation}
    \tilde{E} = E + s_{||}~\Delta E + s_{\perp}~\rl{\Delta E_\perp}~.
\end{equation}
The new energy $\tilde{E}$ is shifted away from the MPTD energy with a parallel and a perpendicular shift. In fact, all the quantities in \eqref{eqn: tilde-conserved-quantities} have a similar structure. Since we want to work at $O(s)$ and we have already factored out the spin dependence, it is sufficient to evaluate the shifts at geodesic order. Because the coefficients $\alpha$ and $\beta$ are defined with respect to these parallel and perpendicular pieces individually, the parallel and perpendicular shifts $\Delta E$ and $\Delta E_\perp$ are functions of $\alpha$ and $\beta$ respectively. 

\subsubsection{Perpendicular shifts}

For the perpendicular spin, one expects no secular effect on the conserved quantities: this is well known after orbit averaging~\cite{WitzHJ.19,Drummond:2022efc,Drummond:2022xej,Pio.25}, and Ref.~\cite{Skoupy:2024uan} showed that it holds locally in Kerr. We find that it holds locally in any KY spacetime, for a suitable choice of $\beta$. The perpendicular shifts of the energy and of the normalisation vanish identically for any $\beta$: $\nabla_a\xi_b$ lies in the plane spanned by $f_{ab}$ and $f^\star_{ab}$, to which $\mathfrak{G}^{ab}$ is orthogonal, while the KY equation makes $\dot\sigma_\perp^a$ orthogonal to both $\xi^a$ and $v^a$ (Appendix~\ref{app:perp-proofs}).

For $\mathcal{L}$ and the Carter constant, by contrast, the perpendicular shifts do not vanish automatically; they reduce to
\begin{subequations}
\begin{align}
    \Delta {L}_\perp &= -\beta~\eta_a\dot{\shift}_\perp^a, \\
    \Delta \GC_\perp &= -2\beta~K_{ab}v^a \dot{\shift}_\perp^b.
\end{align}
\end{subequations}
Therefore, the choice 
\begin{equation}
    \beta = 0
\end{equation}
ensures that the remaining perpendicular shifts also vanish. Explicit calculations for the above expression(s) and proofs for the energy and the normalisation are given in Appendix \ref{app:perp-proofs}. 

\subsubsection{Parallel shifts}
The parallel shifts are all functions of $\alpha$. Unlike the case of perpendicular shifts, these parallel-shifts need not vanish, but must be conserved so that we can associate physical quantities with the auxiliary tangent $\nu$. We find that the shift in the normalisation actually vanishes independent of $\alpha$. For the Carter constant, we find the shift 
\begin{equation}
    \Delta \GC_0 = \fr{-2(\alpha-1)E\GC_0 + 2(3-\alpha){L}}{\sqrt{\GC_0}}~,
\end{equation}
which is conserved for any constant value of $\alpha$ (for an explicit calculation of $\Delta \GC_0$ and $\Delta N$ see Appendix \ref{app:par-proofs}). 
Meanwhile, the conservation of the shift in energy and angular momentum is more subtle. Computing time derivatives, we find (see Appendix \ref{app:par-proofs} for details)
\begin{subequations}\label{eqn:dE-dL-cons}
    \begin{align}
    \dot{\left(\Delta E\right)} =  v^a\nabla_a\left(\Delta E\right) &= \frac{\left(3-2\alpha\right)}{\sqrt{\GC_0}}\xi_a\dot{\xi}^a + \frac{1}{\sqrt{\GC_0}} v^cf^\star_{ab} R^a_c\xi^b~, \\ 
    \dot{\left(\Delta {L}\right)} = v^a\nabla_a\left(\Delta {L}\right) &= \frac{\left(2\alpha-3\right)}{\sqrt{\GC_0}}\eta_a\dot{\xi}^a - \frac{1}{\sqrt{\GC_0}} v^c f^\star_{ab}R^a_c\eta^b~, 
    \end{align}
\end{subequations}
where $R_{ab}$ is the Ricci tensor. 

The choice
\begin{equation}
    \alpha = {3}/{2}
\end{equation}
removes the first term in each line of \eqref{eqn:dE-dL-cons}. In Ricci-flat spacetimes, the parallel shifts are then conserved, and this choice reproduces the auxiliary vector of Ref.~\cite{Skoupy:2024uan} in Kerr. In general, however, the Ricci term survives, and no choice of $\alpha$ removes it. It is nonetheless a total derivative: defining
\begin{equation}\label{eq:def-Psi}
    \Psi_\xi := f_\star^{ab}\xi_{a;b}-\xi_a \xi^a
    \quand
    \Psi_\eta := f^{ab}_\star \eta_{a;b}-\xi_a\eta^a,
\end{equation}
and using \eqref{eq:KY-dxi} and \eqref{eq:KY-deta}, one finds (see Appendix~\ref{app:ids} for derivation)
\begin{equation}\label{eqn:pot-origins}
   f^\star_{ab}R^a_{\phantom{a}c}\xi^b = \frac{1}{2}\nabla_c\Psi_\xi
   \quand
   f^\star_{ab}R^a_{\phantom{a}c}\eta^b = \frac{1}{2}\nabla_c\Psi_\eta .
\end{equation}
Hence, for $\alpha = 3/2$, the combinations
\begin{equation}
    \Delta E - \frac{\Psi_\xi}{2\sqrt{\GC_0}}
    \quand
    \Delta\mathcal{L} + \frac{\Psi_\eta}{2\sqrt{\GC_0}}
\end{equation}
are conserved, although the parallel shifts themselves are not.

In terms of the fiducial tangent $\nu^a$, we have therefore generalised conserved quantities
\begin{subequations}\label{eqn:script-E-L}
\begin{align}
    \mathscr{E} & := E+ s_{||}\left[\Delta E-\fr{1}{2\sqrt{\GC_0}} \Psi_\xi\right] = -\nu_a\tilde{\xi}^a - \fr{s_{||}}{2\sqrt{\GC_0}} \Psi_\xi~, \\
    \mathscr{L} & := {L} + s_{||}\left[\Delta {L} + \fr{1}{2\sqrt{\GC_0}}\Psi_\eta  \right] = \nu_a\tilde{\eta}^a + \fr{s_{||}}{2\sqrt{\GC_0}}\Psi_\eta~.
\end{align}
\end{subequations}
These are the conserved quantities of a particle with effective charge $s_\parallel/\sqrt{\GC_0}$, coupled to a potential whose projections onto $\xi$ and $\eta$ are $\Psi_\xi/2$ and $\Psi_\eta/2$. We make this correspondence precise in Sec.~\ref{sec:eff-eom}.

\section{From spin to charge: effective motion and separability} 
\label{sec:eff-eom}

We summarize our finding in Sec.~\ref{sec: The-New-Worldline} by stating that, at $O(s)$, the shifted worldline of a fiducial orbit admits four first integrals in terms of the tangent $\nu^a$: 
\begin{subequations}\label{eqn:Gen-Shifted-Constants}
\begin{align}
  \mathscr{E}&=-\nu_\tsf{a}\tilde{\xi}^\tsf{a}-\frac{q}{2}\Psi_\xi = E - \fr{q}{2}E^2, \\ 
  \mathscr{L}&=\nu_\tsf{a}\tilde{\eta}^\tsf{a}+\frac{q}{2}\Psi_\eta = {L} - \fr{q}{2}E {L} ,\\
  \widetilde \GC_0&=\tilde{K}_{\tsf{ab}}\nu^\tsf{a}\nu^\tsf{b} = \GC_0 - q\left(3{L}+E\GC_0\right), \phantom{\frac12} \\
   -1&=\nu_\tsf{a}\nu^\tsf{a},
\end{align}
\end{subequations}
where the effective charge is (with $K$ the R\"udiger invariant in~\eqref{eq:KY-Rudiger},)
\begin{equation}\label{eq:def-q}
    q := \frac{s_\parallel}{\sqrt{\GC_0}} = \frac{K}{\GC_0}\,,
\end{equation}
and the potentials $\Psi_{\xi,\eta}$ are defined in \eqref{eq:def-Psi}. These are precisely the forms one expects for a charged particle in a spacetime with KY symmetry, such as Kerr-Newman spacetime. 

This is no coincidence. In this section, we show that spin can play the role of an effective charge. We first identify an effective gauge field $A$ whose field strength $F$ is determined by the principal tensor $ f^\star = h$ and the Ricci tensor $R$; the field vanishes in Ricci-flat spacetimes, recovering the familiar geodesic description in Kerr~\cite{Skoupy:2024uan}, while nonzero Ricci curvature generally induces a Lorentz force. We then work in the off-shell canonical metric with the canonical coordinate $(\ostime, r, y, \varphi)$
\footnote{Note that $\ostime$ denotes a Killing coordinate, which is sometimes denoted by $\pt$ in the literature (and reserved for the proper time here).},
which locally describes the regular class of spacetimes admitting a principal tensor, and show that the equations of motion separate in $r$ and $y$ for arbitrary metric functions $\Delta(r)$ and $\Theta(y)$. 
The resulting solutions also satisfy the Lorentz-force equation at linear order in spin.  

Strictly speaking, this equation governs the motion described by $\nu^a$, rather than the shifted worldline $({\tilde x}^a, {\tilde u}^a)$ itself. Nevertheless, the relation between the two is simple: their projections onto $(r,y,\varphi)$ coincide, up to a constant rescaling of the evolution parameter, while their $\ostime$ coordinates differ by an additional drift. Thus, the dynamics of a neutral spinning body can be represented, through the worldline-shift construction, in terms of an auxiliary non-spinning charged-particle motion. Spin plays the role of an effective charge not merely in the first integrals, but in the equations of motion themselves. 
We illustrate this correspondence in Kerr--de Sitter spacetime in Sec.~\ref{sec:Kerr-dS example}, where $F\propto\Lambda$.

\subsection{An effective gauge field from the principal tensor}
\label{subsec:gauge}

The first integrals \eqref{eqn:Gen-Shifted-Constants} suggest that $\Psi_\xi/2$ and $\Psi_\eta/2$ play the role of the projections $A_a\xi^a$ and $A_a\eta^a$ of a gauge potential coupled to the charge $q$. To identify the underlying field, we compute the gradients of the potentials. Using \eqref{eq:KY-dxi}, one finds
\begin{equation}\label{eq:dPsi-F}
    \nabla_a\Psi_\xi = 2f^\star_{ab}R^{bc}\,\xi_c,
    \quand
    \nabla_a\Psi_\eta = 2f^\star_{ab}R^{bc}\,\eta_c.
\end{equation}
Both are contractions of the same antisymmetric tensor with a Killing vector. This singles out the natural candidate
\begin{equation}\label{eq:def-F}
    F_{ab} := f^\star_{ac}R_b{}^c ,
\end{equation}
built solely from the principal tensor $f^\star_{ab}$ and the Ricci tensor $R_{ab}$.
Heuristically, the shift aligns the parallel spin tensor with the principal tensor,
$s_\parallel\mathfrak F^{ab}=-q\,f^{ab}_\star $, so that the parallel spin--curvature coupling
splits, via \eqref{eq:KY-dxi}, into a term proportional to $\dot\xi^a$ and the Ricci term
$qF^a{}_b\nu^b$. For $\alpha=3/2$ the $\dot\xi^a$ terms cancel exactly, and the perpendicular
contribution vanishes for $\beta=0$. A covariant derivation is given in
Appendix~\ref{app:cov-proof}; the explicit solution~\eqref{eq:nu-sep} was also checked to
satisfy~\eqref{eq:Lorentz-nu} in the off-shell metric.

This argument, however, is only suggestive, and Eq.~\eqref{eq:dPsi-F} fixes merely the contractions of $F$ with $\xi$ and $\eta$. We therefore take \eqref{eq:def-F} as a definition and verify that it is a legitimate field strength. First, $F$ is antisymmetric by \eqref{eq:KY-dxi}. Second, $F$ is closed, $\nabla_{[a}F_{bc]}=0$: evaluating $\nabla_aF_{bc}$ with \eqref{eq:KY-dxi} and Kostant's formula for the second derivative of $\xi$, and antisymmetrizing over $[abc]$, all terms cancel by the Bianchi identities. Hence, locally, $F=\mathrm{d}A$ for some gauge potential $A$, sourced by the current
\begin{equation}\label{eq:def-J}
    \mc{J}^a := \nabla_bF^{ab}
    = R^{ab}\xi_b + \frac{1}{2} R \,\xi^a
      + \frac{1}{2}\nabla_c\!\left(Rf_\star^{ac}\right).
\end{equation}
Both $F^{ab}$ and $\mc{J}^a$ vanish in Ricci-flat spacetimes, so the gauge field is a genuinely non-vacuum feature. We stress that $F^{ab}$ is not a Maxwell field: it is sourced by the spacetime curvature rather than by an electromagnetic current. In the Kerr--Newman spacetime, for instance, the metric depends on the (Newman) charge parameter $\frak{q}$ only through $\frak{q}^2$, so $F^{ab}$ is necessarily even in it. Explicitly, $F_{r\zeta} = \frak{q}^2 r\Sigma^{-2}$, whereas the Maxwell field is linear in $\frak{q}$.

The potentials $\Psi_{\xi,\eta}$ are invariant along both Killing vectors, since they are built from $g_{ab}$, $f^\star_{ab}$, $\xi^a$, and $\eta^a$, all of which are preserved by $\xi$ and $\eta$ (the two Killing vectors commute and preserve the principal tensor $f^\star_{ab}$ --- see e.g. Appendix D in~\cite{10.1093/acprof:oso/9780199692293.001.0001}).
By \eqref{eq:dPsi-F}, this is equivalent to $F(\xi,\eta)=0$. We may
therefore choose a gauge in which
$\mathcal{L}_\xi A = \mathcal{L}_\eta A = 0$.
Cartan's formula then gives $\nabla_a(A_b\xi^b) = F_{ab}\xi^b$, and likewise for $\eta$.
Comparing with \eqref{eq:dPsi-F}, we may set
\begin{equation}\label{eq:A-proj}
    A_a\xi^a = \frac{\Psi_\xi}{2},
    \qquad
    A_a\eta^a = \frac{\Psi_\eta}{2},
\end{equation}
where additive constants have been absorbed into the gauge. The first
integrals \eqref{eqn:Gen-Shifted-Constants} then take exactly the form
\begin{equation}
    \mathscr{E} = -\nu_a\xi^a - qA_a\xi^a,
    \qquad
    \mathscr{L} = \nu_a\eta^a + qA_a\eta^a,
\end{equation}
familiar from a particle of charge $q$ coupled to $A^a$ \cite{Ha.15}. 

\subsection{Separable motion in the off-shell metric}
\label{subsec:sep}

Whether the motion itself is governed by the corresponding Lorentz force is a separate question. To answer it, we need the motion explicitly in a given coordinate system, and for this we exploit the strong constraints that a principal tensor $f^\star_{ab}$ imposes on the geometry in Sec.~\ref{sec:KY}. 
The idea is simple. In Kerr spacetime (with the spin parameter $a$), the eigenvalues of $f^\star_{ab}$ are $r$ and $a\cos\theta$, and together with $t$ and $\phi$ they form the familiar Boyer--Lindquist coordinates. More generally, the eigenvalues $(r,y)$ of $f^\star_{ab}$, which are defined by the same eigenplanes as the Darboux frame of Sec.~\ref{sec:Marck}, can serve as coordinates. Supplemented by Killing-vector direction $(\ostime,\varphi)$ adapted to $\xi$ and $\eta$, they put the metric into a universal form.

We work in a regular local domain where the eigenvalues $r$ and $y$ of the principal tensor are functionally independent, $\mathrm{d}r \wedge \mathrm{d}y\neq0$, and have non-null gradients, so that they can serve as coordinates. We further assume that the associated Killing vectors $\xi$ and $\eta$ are linearly independent there~\cite{Krtous:2008tb,Houri:2008ng,Frolov:2017kze}.
These regularity assumptions exclude the exceptional null-eigenvalue classes possible in Lorentzian signature~\cite{Frolov:2017whj}. In such a domain, the metric and its inverse locally takes the off-shell canonical form (see, e.g., Sec.~3 of Ref.~\cite{Frolov:2017kze}),
\begin{subequations}
\begin{align}
  g &= -\frac{\Delta}{\Sigma}\,\omega_r^2 + \frac{\Theta}{\Sigma}\,\omega_y^2
       + \frac{\Sigma}{\Delta}\,\mathrm{d}r^2 + \frac{\Sigma}{\Theta}\,\mathrm{d}y^2,
  \label{eq:off-shell}\\
  \Sigma\, g^{-1} &= \Delta\,\partial_r^2 + \Theta\,\partial_y^2
       - \frac{V_r^2}{\Delta} + \frac{V_y^2}{\Theta},
  \label{eq:off-shell-inv}
\end{align}
\end{subequations}
where $\Sigma = r^2+y^2$ and
\begin{equation}\label{eq:omega-V}
  \omega_r = \mathrm{d}\ostime + y^2\,\mathrm{d}\varphi, \quad
  \omega_y = \mathrm{d}\ostime - r^2\,\mathrm{d}\varphi, \quad
  V_r = r^2\partial_\ostime + \partial_\varphi, \quand
  V_y = y^2\partial_\ostime - \partial_\varphi .
\end{equation}
Here $\Delta(r)$ and $\Theta(y)$ are arbitrary functions; no field equations have been imposed, hence the name ``off-shell.'' Specific choices of $\Delta$ and $\Theta$ reproduce the familiar
Kerr--Newman--(A)dS--NUT family. The form is regular away from $\Delta=0$, $\Theta=0$, and $\Sigma=0$. In particular, $r$ and $y$ enter only through $\Sigma$, so the surfaces $r=0$ and $y=0$ are not excluded individually. In Kerr, this means that the equatorial plane $y=0$ is covered and only the ring $r = 0 = y$ is removed.

The vectors $V_r/\Sigma$ and $V_y/\Sigma$ in \eqref{eq:omega-V} are dual to $\omega_r$ and $\omega_y$. The metric \eqref{eq:off-shell} admits the principal tensor and the KY tensor
\begin{equation}
  f^\star = y\,\mathrm{d}y\wedge\omega_y - r\,\mathrm{d}r\wedge\omega_r,
  \quand
  f = r\,\mathrm{d}y\wedge\omega_y + y\,\mathrm{d}r\wedge\omega_r ,
\end{equation}
which generate the Killing-St\"ackel tensor $K_{ab}$ and two Killing vectors: $\xi = \partial_\ostime$ and $\eta^a = -\partial_\varphi$ (cf. Sec.~\ref{sec:KY}).
By this duality, the gauge potential $A_a$ of Sec.~\ref{subsec:gauge} decomposes as
\begin{equation}\label{eq:A-sep}
  A = \frac{e(r)\,\omega_r + h(y)\,\omega_y}{\Sigma},
\end{equation}
with
\begin{equation}\label{eq:def-eg}
    e(r) := A(V_r) = \frac{1}{2}\left(r\Delta' - \Delta\right),
    \quand
    h(y) := A(V_y) = \frac{1}{2}\left(\Theta - y\,\Theta'\right),
\end{equation}
where primes denote derivatives with respect to the argument.
Each coefficient depends on a single coordinate, and $A$ is in
Lorenz gauge, $\nabla^a A_a = 0$.

We now solve the first integrals \eqref{eqn:Gen-Shifted-Constants} for $\nu^a$, closely following Carter's treatment of charged particles in the Kerr--Newman spacetime~\cite{Carter.68}. 
Introducing the auxiliary functions
\begin{equation}\label{eq:X-functions}
    X_r(r) := -r^2\mathscr{E} - \mathscr{L} - q\,e(r),
    \quand
    X_y(y) := -y^2\mathscr{E} + \mathscr{L} - q\,g(y),
\end{equation}
we find
\begin{subequations}\label{eq:nu-sep}
\begin{align}
  \Sigma\,\nu^\ostime
    &= -\frac{r^2X_r}{\Delta} + \frac{y^2X_y}{\Theta}, &
  \Sigma\,\nu^\varphi
    &= -\frac{X_r}{\Delta} - \frac{X_y}{\Theta}, \\
  \left(\Sigma\,\nu^r\right)^2
    &= X_r^2 - \Delta\,(r^2 + \widetilde{\GC}_0), &
  \left(\Sigma\,\nu^y\right)^2
    &= \Theta\,(\widetilde{\GC}_0 - y^2) - X_y^2 .
\end{align}
\end{subequations}
Each component depends on $r$ and $y$ only through $X_r$, $\Delta$ and $X_y$, $\Theta$, respectively, apart from the overall factor $\Sigma$.
Substituting \eqref{eq:nu-sep} back into \eqref{eqn:Gen-Shifted-Constants} reproduces all four first integrals.

More importantly, the vector $\nu^a$ obeys the Lorentz-force equation,
\begin{equation}\label{eq:Lorentz-nu}
    \nu^b\, \nabla_b\, \nu^a = q\,F^a{}_b\,\nu^b + O(s^2).
\end{equation}
We verified this by evaluating both sides in the off-shell metric with $\nu^a$ given by \eqref{eq:nu-sep}, keeping $\Delta(r)$ and $\Theta(y)$ arbitrary.
Thus, not only the first integrals but the equation of motion itself is that of a particle of charge $q$ in the field $F$. In Ricci-flat spacetimes, $F=0$ and $\nu^a$ is a geodesic, as is the case in the Kerr geometry~\cite{Skoupy:2024uan}; otherwise, the motion is forced. Since the off-shell metric locally describes every spacetime in the class considered here~\cite{Frolov:2017kze}, the equations of motion \eqref{eq:Lorentz-nu} hold throughout that class. 
A manifestly covariant derivation is given in Appendix~\ref{app:cov-proof}. It also shows that the Lorentz-force form in ~\eqref{eq:Lorentz-nu} holds only for $\alpha = 3/2$: the choice made in Sec.~\ref{sec: The-New-Worldline} to obtain conserved potentials $\Psi_{\xi,\eta}$ is thus singled out independently by the equation of motion.

We now return to the shifted worldline. From \eqref{eq:nu-ansatz} and
\eqref{eq:derdx3}, its tangent is
\begin{equation}\label{eq:tilde-u-nu}
    \tilde u^\tsf{a} = \frac{\mathrm{d}\tilde x^\tsf{a}}{\mathrm{d}\pt}
    = \left(1 + \frac{3 s_\parallel E}{2\sqrt{\GC_0}}\right)\nu^\tsf{a}
      - \frac{3 s_\parallel}{2\sqrt{\GC_0}}\,\tilde{\xi}^\tsf{a} .
\end{equation}
Since $\xi = \partial_\ostime$, the shifted worldline differs from the
auxiliary curve only by a constant rescaling of its parameter and a
drift in $\ostime$. 

Both are absorbed by the deformed Carter--Mino time $\tilde\lambda$
introduced in Ref.~\cite{Skoupy:2024uan} for the Kerr spacetime,
\begin{equation}\label{eq:deformed-Mino}
    \frac{\mathrm{d}\pt}{\mathrm{d}\tilde\lambda}
    = \tilde\Sigma\left(1 - \frac{3 s_\parallel E}{2\sqrt{\GC_0}}\right),
\end{equation}
where $\tilde\Sigma$ is $\Sigma$ evaluated along the shifted worldline.
Since the rescaling factor involves only $E$, $\GC_0$, and $s_\parallel$, this definition carries over unchanged to the off-shell metric.
To linear order in spin, we obtain
\begin{subequations}\label{eqn:Gen-Sep-Sols}
\begin{align}
    \frac{\mathrm{d}\tilde\ostime}{\mathrm{d}\tilde\lambda}
      &= \frac{y^2X_y}{\Theta} - \frac{r^2X_r}{\Delta}
         - \frac{3 s_\parallel}{2\sqrt{\GC_0}}\,\tilde\Sigma, &
    \frac{\mathrm{d}\tilde\varphi}{\mathrm{d}\tilde\lambda}
      &= -\frac{X_r}{\Delta} - \frac{X_y}{\Theta}, \\
    \left(\frac{\mathrm{d}\tilde r}{\mathrm{d}\tilde\lambda}\right)^2
      &= X_r^2 - \Delta\,(r^2 + \widetilde{\GC}_0), &
    \left(\frac{\mathrm{d}\tilde y}{\mathrm{d}\tilde\lambda}\right)^2
      &= \Theta\,(\widetilde{\GC}_0 - y^2) - X_y^2 .
\end{align}
\end{subequations}
The $\varphi$, $r$, and $y$ equations coincide with those of $\nu^a$ in \eqref{eq:nu-sep}. The shifted worldline therefore traces the same path in $(r,y,\varphi)$ and inherits its separability; the only difference is the last term in the $\ostime$ equation. A term of the same form also appears in Kerr and can be treated as in Ref.~\cite{Skoupy:2024uan}.

To reconstruct the physical MPTD solution, we also need the Marck phase $\psi$ discussed in Sec.~\ref{sec:Marck}: it fixes the orientation of the spin~\eqref{eqn: spin-vect-Marck} and enters the shift itself through $\shift^a$, so that $x^a = \tilde x^a - \shift^a$ can be recovered with it. Since $\psi$ always appears multiplied by $s_\perp$, it is needed only at zeroth order in spin, i.e., along the background geodesic.
Along the background geodesic, it obeys a separable equation, as is known for spacetimes admitting a (non-degenerate) KY tensor~\cite{KamMar.86}; in the off-shell metric it reads
\begin{equation}\label{eq:Marck-angle-sep}
    \frac{\mathrm{d}\psi}{\mathrm{d}\lambda}
    = \sqrt{\GC_0}\,\bigl(\mathcal{L} - \GC_0 E\bigr)
      \left(\frac{1}{\GC_0 - y^2} - \frac{1}{r^2 + \GC_0}\right),
\end{equation}
where $\lambda$ is the Carter--Mino time. Notably, this equation does not involve $\Delta$ or $\Theta$. To the order retained here, $r$ and $y$ on the right-hand side may be evaluated along the shifted orbit, with $\lambda \to \tilde\lambda$ at the cost of an $O(s^2)$ error.
Together with~\eqref{eqn:Gen-Sep-Sols}, this completes the construction: for any $\Delta(r)$ and $\Theta(y)$, the motion of the spinning body at $O(s)$, including its spin orientation, reduces to separated quadratures.

\subsection{Example: Kerr--de Sitter spacetime}
\label{sec:Kerr-dS example}

As an illustration, we consider the Kerr--de Sitter spacetime with mass parameter $M$, spin parameter $a\neq0$, and cosmological constant $\Lambda\geq0$
In Boyer--Lindquist coordinates $(t,r,\theta,\phi)$, its metric takes the standard form (e.g., see the review~\cite{Frolov:2017kze}). 
It follows from the off-shell metric \eqref{eq:off-shell} via
\footnote{
The factor $1/a$ in the transformation reflects the fact that the Schwarzschild--de Sitter limit $a \to 0$ lies outside the regular domain assumed above. Because $y=az$ vanishes identically, $\mathrm{d}r \wedge \mathrm{d}y=0$ and the eigenvalues of the principal tensor no longer serve as coordinates. This is a degeneracy of the canonical coordinates, not of the physics. Indeed, once $L$ is expressed in terms of $J_z$, Eqs.~\eqref{eq:Kerr-dS-sep-eqns} admit a smooth limit $a \to 0$; a dedicated treatment of the spherically symmetric case is nonetheless left to the literature (e.g. Refs.~\cite{TheocharisAApostolatos_1996,WiPio.23,Ra.Iso.Dru.IntegO2.26}).}
%
\begin{equation}
    y = az, \qquad
    \ostime = t - \frac{a}{\chi}\,\phi, \qquad
    \varphi = \frac{\phi}{a\chi},
\end{equation}
with $z=\cos\theta$ and $\chi = 1+\Lambda a^2/3$, and our choice is 
\begin{equation}
    \Delta(r) = (r^2+a^2)\left(1-\frac{\Lambda r^2}{3}\right) - 2Mr,
    \qquad
    \Theta(z) = a^2(1-z^2)\left(1+\frac{\Lambda a^2z^2}{3}\right),
\end{equation}
so that $\Sigma = r^2 + a^2 z^2$. The factor $\chi$ ensures that $\phi$ has period $2\pi$, avoiding a conical singularity on the axis. 

The Kerr--de Sitter spacetime is an Einstein space, $R_{ab} = \Lambda g_{ab}$. The general results of Sec.~\ref{subsec:gauge} then simplify considerably. The field strength is proportional to the principal tensor $f^\star_{ab}$, and the current to the Killing vector,
\begin{equation}\label{eq:KdS-F}
    F = \Lambda f^\star, \qquad \mathcal{J} = -3\Lambda\,\xi .
\end{equation}
Both are sourced by the cosmological constant and vanish in the Kerr limit. 
In Boyer--Lindquist coordinates, the corresponding potential is
\begin{equation}\label{eq:KdS-A}
    A = \frac{1}{2}\left(1 - \frac{\Lambda a^2}{3} - \Lambda r^2
        + \Lambda a^2 z^2\right)\mathrm{d}t
      - \frac{a}{2\chi}\left[2 - \frac{\Lambda a^2}{3}
        - \Lambda r^2(1-z^2) + \Lambda a^2 z^2\right]\mathrm{d}\phi ,
\end{equation}
which reduces to the pure gauge $ A = \tfrac{1}{2}\mathrm{d}t - a\,\mathrm{d}\phi$ as $\Lambda \to 0$.

The functions \eqref{eq:X-functions} follow from $e(r)$ and $h(z)$ in Eq.~\eqref{eq:def-eg}.
Explicitly,
\begin{subequations}
\begin{align}
    X_r &= -\left(\mathscr{E}+\frac{q}{2} \right) r^2 
            - \mathscr{L} + \frac{q}{2}a^2
            + \frac{q}{2}\Lambda r^2 \left(r^2+\frac{a^2}{3} \right), \\
    X_z &= -\left( \mathscr{E}+\frac{q}{2} \right)a^2z^2 
            + \mathscr{L} - \frac{q}{2}a^2
            - \frac{q}{2}\Lambda a^4z^2 \left(z^2-\frac{1}{3} \right).
\end{align}
\end{subequations}
and the shifted worldline obeys
\begin{subequations}\label{eq:Kerr-dS-sep-eqns}
\begin{align}
    \frac{\mathrm{d}\tilde t}{\mathrm{d}\tilde\lambda}
      &= -\frac{X_z}{1+\Lambda a^2z^2/3}
         - \frac{(r^2+a^2)\,X_r}{\Delta}
         - \frac{3s_\parallel}{2\sqrt{\GC_0}}\,\tilde\Sigma, \\
    \frac{\mathrm{d}\tilde\phi}{\mathrm{d}\tilde\lambda}
      &= -a\chi\left(\frac{X_z}{\Theta} + \frac{X_r}{\Delta}\right), \\
    \left(\frac{\mathrm{d}\tilde r}{\mathrm{d}\tilde\lambda}\right)^2
      &= X_r^2 - \Delta\,(r^2 + \widetilde{\GC}_0)\, \\ 
    a^2\,\left(\frac{\mathrm{d}\tilde z}{\mathrm{d}\tilde\lambda}\right)^2
      &= {\Theta\,(\widetilde{\GC}_0 - a^2z^2) - X_z^2}. 
\end{align}
\end{subequations}
Here $\mathcal{L}$ is defined with respect to $\eta$, not $\partial_\phi$. It is related to the usual azimuthal angular momentum by $J_z = (a^2E - \mathcal{L})/(a\chi)$.

For $\Lambda=0$, the charge enters only through constant shifts $\mathscr{E} \to \mathscr{E} + q / 2$ and $\mathscr{L} \to \mathscr{L} \mp q a^2/2$, which is a pure-gauge effect, and the equations reduce to the Kerr results of Ref.~\cite{Skoupy:2024uan}: geodesic equations with shifted constants of motion. 
For $\Lambda>0$, the last terms in $X_r$ and $X_z$ carry the genuine coupling to the field $F$ in ~\eqref{eq:KdS-F}, yet the equations remain separated in $r$ and $z$; they have the same functional form as those of a charged particle in the Kerr--Newman spacetime~\cite{Carter.68}, with the Maxwell field replaced by $F= \Lambda\, f^\star$.

\section{Discussion and conclusions}
\label{sec:discussion}

In this work, we showed that the worldline-shift construction proposed for Kerr spacetime~\cite{Skoupy:2024uan} extends, covariantly, to any spacetime admitting a non-degenerate Killing--Yano tensor, without assuming any field equations. Spin, at linear order, then plays the role of charge: after the shift, the motion is that of a fictitious spinless body of charge $q = s_\parallel/\sqrt{\GC_0}$, obeying the Lorentz-force equation~\eqref{eq:Lorentz-nu} with an effective field strength built from the principal tensor and the Ricci tensor. In the off-shell metric~\cite{Frolov:2017kze}, the resulting equations of motion separate, both for the auxiliary curve, Eq.~\eqref{eq:nu-sep}, and for the shifted worldline, Eq.~\eqref{eqn:Gen-Sep-Sols}. The field vanishes in Ricci-flat spacetimes, recovering the geodesic description in Kerr, whereas in Kerr--de Sitter it is simply $F=\Lambda f^\star$ (Sec.~\ref{sec:Kerr-dS example}). Since the construction does not depend on the nature of the orbit, it applies to bound and unbound motion alike.

The integrability of the linear-in-spin system is known~\cite{Ra.Iso.IntegO1.26,Ra.Iso.PapII.24,Ra.CQG.24,Witzany_2025,Witzany:2019nml}; our construction realises it explicitly (see also Ref.~\cite{Kubiznak:2011ay} for a related model). The hidden symmetry reviewed in Sec.~\ref{sec:intro-hidden} is thus not only sufficient for integrability, but also, at linear order in spin, for separable equations of motion. For the modelling of small-mass-ratio binaries at first post-adiabatic order, these solutions may serve as input to waveform-generation pipelines, allowing linear-in-spin effects of the secondary to be tracked consistently in spacetimes beyond Kerr.

Several directions remain open. As recalled in Sec.~\ref{sec:intro}, integrability at quadratic order in spin holds only for black holes~\cite{Ra.Iso.Dru.IntegO2.26,Ra.NoCarter.26}; whether the Killing--Yano tensor also yields separable equations of motion at that order is an open question. The linear-in-spin solutions could also be used to study secondary-spin effects on fluxes and on resonances in spacetimes beyond Kerr. Finally, the Kerr--NUT--(A)dS metrics in higher dimensions also admit a principal tensor~\cite{Krtou08,Frolov:2017kze}, and their geodesic motion is integrable~\cite{PKVK.07}. It would be interesting to ask whether spin breaks this integrability and, if not, how far the Killing--Yano structure alone can carry the construction presented here.

\begin{acknowledgments}
We thank Abraham Harte, Josh Mathews, Viktor Skoup\'{y} and Vojt\v{e}ch Witzany for useful discussions. S.I.~also thanks Alvin J. K.~Chua for his continuous encouragement.
P.R.~acknowledges funding from the France 2030 program, as part of the 3NC project. S.I.~is supported by the Ministry of Education, Singapore, under the Academic Research Fund Tier 1 A-8001492-00-00 (FY2023) and A-8004174-00-00 (FY2025). S.R.D.~acknowledges financial support from the Science and Technology Facilities Council (STFC) under Grant No. ST/X000621/1. A.N.S.~is supported by an EPSRC PhD Studentship.  

The scientific content and derivations are the authors' own; AI tools were used to assist with coding and with editing the manuscript text, and all output was checked by the authors. A.N.S used Claude Sonnet 5.5 for background reading and research. S.I. used Claude Opus 5.5 and GPT 6 to assist with coding, and manuscript editing.
\end{acknowledgments}

\bibliography{draft_ref.bib,ListRef.bib}

\appendix

\section{Conventions and notations}
\label{app:conventions}

Throughout the paper, we use geometric units in which $G=c=1$, and the metric signature is $(-,+,+,+)$.
Abstract indices are denoted by $a,b,c,\dots$. For components, we use two kinds of indices ranging over $\{0,1,2,3\}$: Greek indices $\alpha,\beta,\gamma,\dots$ label components in the natural basis $(\partial_\alpha)^a$ associated with a spacetime coordinate system $x^\alpha$,
and capital Latin indices $A,B,C,\dots$ label components in an orthonormal tetrad basis $(e_A)^a$.

Symmetrization and anti-symmetrization are defined by
$T_{(ab)} := \tfrac{1}{2}(T_{ab}+T_{ba})$ and
$T_{[ab]} := \tfrac{1}{2}(T_{ab}-T_{ba})$, respectively.
Our curvature conventions follow Wald's textbook \cite{Wald}, in particular, $\nabla_{[a}\nabla_{b]}\omega_c = \tfrac{1}{2}\,R_{abc}{}^{d}\,\omega_d$ for any $1$-form $\omega_a$.

We take $\varepsilon_{abcd}$ to be the volume form compatible with the metric, which is totally anti-symmetric and satisfies $\varepsilon_{0123}=+\sqrt{-g}$ in any right-handed coordinate system. In an orthonormal tetrad, $\varepsilon_{0123}=+1$. 
We use the term ``Killing--Yano spacetime'' as shorthand for a spacetime endowed with a non-degenerate Killing--Yano tensor, i.e., an anti-symmetric Killing--Yano $2$-form $f_{ab}$ of matrix rank $4$. We define its Hodge dual as $f^\star_{ab} = \varepsilon_{ab}~^{cd}f_{cd}/2$, using $f_\star$ or $f^\star$ interchangeably. Given some tensor $A_{ab}$ on $\mathscr{W}$, when evaluated on the shifted worldline we denote this object as $\tilde{A}_\tsf{ab}$ with indices in \tsf{sans serif} text. We also use semi-colons (commas) to denote covariant (partial) derivatives i.e. $A_{a;b} = \nabla_b A_a$ and $A_{a,b} = \partial_b A_a$. 

\section{Bitensors and parallel propagators}

\label{app:shift-setup}

In this appendix we collect some background material on bitensors from Ref.~\cite{Poisson:2011nh}, with the intention to make this paper reasonably self--contained. The general theory of bitensors is detailed in Part. I of the review by Poisson, Pound and Vega~\cite{Poisson:2011nh}, which itself is based on Synge’s book~\cite{Synge1960} and the work by DeWitt and Brehme~\cite{DeWitt:1960fc}. 

We consider bitensors between the source point $x$ and the field point $\tilde{x}$.  Tensorial objects at these points have indices $a,b,c$ and $\tsf{a}, \tsf{b}, \tsf{c}$ respectively i.e. we will use \tsf{sans serif} alphabets to denote tensors evaluated at the field point(s). Assuming that $x$ is connected to $\tilde{x}$ through a unique geodesic $\beta$ (i.e. $\tilde{x} \in \mc{N}(x)$) we parametrise this geodesic with $z(\lambda)$ where $\lambda$ is \emph{some} affine parameter along the geodesic such that $z(\lambda_0) = x$ and $z(\lambda_1) = \tilde{x}$. Arbitrary points on a geodesic will be labelled by the indices $a,b,c,...$ or $\tsf{a,b,c,...}$. 

In this framework, Synge's world function generalises the notion of coordinate displacement and is covariantly defined as, 
\begin{equation}
    \sigma(\tilde{x},x) 
    = \frac{\left(\Delta \lambda\right)^2}{2}\varepsilon~, 
\end{equation}
where $g_{ab}\dot{z}^a \dot{z}^b = \varepsilon$ is the normalisation of the geodesic tangent. $\varepsilon \in \{0, \pm 1\}$ depending on whether $\beta$ is null, spacelike or timelike respectively. One can show that first derivatives of the Synge world function are proportional to the geodesic tangents i.e. $\sigma^\tsf{a} = \Delta \lambda \dot{z}^\tsf{a}$ and $\sigma^{a}= -\Delta \lambda \dot{z}^{a}$. 

On $\beta$ construct a tetrad $(e_A)^a$ where $A = 0,1,2,3$ labels the tetrad legs, which is parallel transported along the geodesic i.e. 
\begin{equation}
    g_{ab}(e_A)^a (e_B)^b = \eta_{AB}~\quand~\frac{D (e_A)^a}{\ud\lambda} = 0~.
\end{equation}

We denote by $(e^A)_a$ the 1-forms such that $(e^A)_a=\eta^{AB}g_{ab}(e_B)^b$, such that $(e^A)_a(e^B)^a=\eta^{AB}$. Now, suppose a vector at $x$ is parallel transported along $\beta$ to the point $\tilde{x}$. In terms of the tetrad components, 
\begin{equation}
    X^{\tsf{a}}(\tilde{x}) = (e_A)^\tsf{a} X^{A} = (e_A)^\tsf{a} (e^{A})_{a} X^{a}(x) := g^\tsf{a}_{~a}(\tilde{x},x)X^{a}(x)~,
\end{equation}
where we have defined the parallel propagator $g^\tsf{a}_{~a}$. The inverse propagator $g^{a}_{~\tsf{a}}$ is defined analogously and parallel transports one-forms. We note that $g^{\tsf{a}}_{~a}g^{a}_{~\tsf{b}} = \delta^\tsf{a}_\tsf{b}$ and $g^{a}_{~\tsf{a}}g^{\tsf{a}}_{~b} = \delta^{a}_{b}$. For an arbitrary tensor we need one copy of the propagator (or its inverse) for each index.

For derivatives of the parallel propagator, since the tetrad legs are parallel transported,
\begin{equation*}
    g^{\tsf{a}}_{~a;\tsf{b}}\sigma^\tsf{b} = \Delta\lambda \dot{z}^\tsf{b} \nabla_\tsf{b}\left[(e_A)^{\tsf{a}} (e^{A})_{a}\right] = 0 = g^{\tsf{a}}_{~a;b}\sigma^{b}~, 
\end{equation*}
with analogous results also for the inverse propagator. Further, $\sigma^\tsf{a} = -g^{\tsf{a}}_{~a}\sigma^{a}$ and $\sigma^{a} = -g^{a}_{~\tsf{a}}\sigma^\tsf{a}$ because $\dot{z}^\tsf{a} = g^\tsf{a}_{~a}\dot{z}^{a}$~.

\section{Covariant Proofs}
\subsection{Deriving the perpendicular shift vector}\label{app:deriv-dxperp}

We want to derive the shift vector $\shift_\perp$ in \eqref{eqn: deriving-dxperp} assuming the result for $\shift_3$, which was derived in \eqref{eqn: deriving-dx3}. To do so, we evaluate the contraction $\mf{G}^{ab}f^\star_{ab}$ which we argue must vanish in the discussion following \eqref{eqn: deriving-dx3}. $\mf{G}^{ab}$ is a complex tensor built out of the legs of the Marck tetrad and the shifted tetrad. Because  the Marck tetrad $\{v^a, (e_A)^a\}$ is right-handed and orthonormal, we can rewrite the Levi-Civita outer products for $\mf{G}^{ab}$ in \eqref{eqn: spin-tensor-split-perp} in terms of the remaining legs. This simplifies the tensor to be 
\begin{equation}\label{eqn: perp-spin-tensor-fancy-G}
    \mf{G}^{ab} = 2 v^{[a}\shift_\perp^{b]} + \varepsilon^{ab}~_{cd}v^{c}e_{\perp}^{d} = 2v^{[a}\shift_\perp^{b]} -2 \ui e_\perp^{[a}(e_3)^{b]}~.
\end{equation}
Now explicitly evaluating the contraction and equating to zero, 
\begin{align}
    0 &= f^\star_{ab}\mf{G}^{ab} = 2f^\star_{ab} \left(v^{a}\shift_\perp^{b} -\ui e_\perp^{a}e_3^{b}\right) = v_a\left(f_\star^{ab}(\shift_\perp)_b + \frac{\ui}{\sqrt{\GC_0}}f^{ba}f^\star_{cb}e_\perp^c\right)~. 
\end{align}
We can see that the contraction of $v^a$ with either term inside the parentheses vanishes by orthogonality. We then rearrange for the vector inside the parentheses to itself vanish, which yields 
\begin{equation*}
    \shift_\perp^a = - \fr{\ui}{\sqrt{\GC_0}}f^a_{\ph b}e_\perp^b~, 
\end{equation*}
as desired. One can argue that the vector in the parentheses may simply be orthogonal to $v$ but not zero, however this simpler choice also renders $\mf{G}$ orthogonal to $f$. Using the expression for $\shift_\perp$ so derived, we show this orthogonality explicitly below
\begin{align}
    f_{ab}\mf{G}^{ab} &= 2f_{ab}\left(v^{a}\shift_\perp^{b} -\ui e_\perp^{a} (e_3)^{b}\right) = \fr{2\ui}{\sqrt{\GC_0}}f_{ab}\left(f^{bc}e_\perp^av_c - f^b_{\ph c} e^c_\perp v^a\right)~, \nonumber \\
    &= -\fr{2\ui}{\sqrt{\GC_0}} \left(K_{ac} e_\perp^a v^c - K_{ac}e_\perp^c v^a\right)= 0~,
\end{align}
where in the second equality we have used the expressions for $\shift_\perp$. Thus, we see that the perpendicular spin tensor $\mf{G}$ is orthogonal to both $f$ and $f^\star$. 

\subsection{Derivatives of the shift vectors}\label{app:derivatives}

In evaluating the parallel and perpendicular shifts to the physical observables we require the time derivatives of the shift vectors. We compute them explicitly here. Firstly, for $\shift_3$, 
\begin{equation}\label{eq:derdx3}
    \dot{\shift}_3^a = -\fr{1}{\sqrt{\GC_0}}f_{\star}^{ab;c}v_bv_c = -\fr{1}{\sqrt{\GC_0}}v_b v_c \left(g^{ac}\xi^b - g^{bc}\xi^a\right) = \fr{1}{\sqrt{\GC_0}}\left(E \,v^a - \xi^a\right) .
\end{equation}
We then automatically see that $\shift_3^a v_a \propto -E + E = 0$. For the (complex) perpendicular vector derived in \eqref{eqn: deriving-dxperp}, firstly note that the vector $e_\perp$ is parallel transported. We then have that the derivative of the perpendicular shift is evaluated as 
\begin{equation}\label{eqn: derdxperp}
    \dot{\shift}_\perp^a = -\fr{\ui}{\sqrt{\GC_0}}f^{a}_{\ph b;c} \, e_\perp^b v^c = -\fr{\ui}{\sqrt{\GC_0}}~\varepsilon^{da}_{\ph\ph bc} \xi_d e_\perp^b v^c = -\fr{\ui}{\sqrt{\GC_0}}\varepsilon^a_{\ph bcd}e_\perp^b\xi^c v^d~,
\end{equation}
using \eqref{eq:KY-xi} in the second equality. Therefore, $\dot{\shift}_\perp$ is perpendicular to both the geodesic velocity $v$ and the primary Killing vector $\xi$. This supports our argument in Sec \ref{sec: The-New-Worldline} that the vector $\tilde{u}$ is not tangent to a physical MPTD worldline. From \eqref{eq:derdx3} and \eqref{eqn: derdxperp}, note that
\begin{equation} \label{ddotshifts}
    \ddot{\shift}_3^a = -\fr{1}{\sqrt{\GC_0}}\dot{\xi}^a~~~\text{and}~~~ \ddot{\shift}_\perp^a = - \fr{i}{\sqrt{\GC_0}}\varepsilon^a_{\ph bcd}e_\perp^b \dot{\xi}^c v^d~. 
\end{equation}
Thus, $\ddot{\shift}_3$ and $\ddot{\shift}_\perp$ are manifestly orthogonal and therefore, the vector $\dot{\shift}^a = s_{||}\dot{\shift}_3^a + s_{\perp} \rl{\dot{\shift}_\perp^a}$ cannot be parallel transported because the individual vectors are not parallel transported. 

\subsection{Vanishing perpendicular shifts to the constants of motion}\label{app:perp-proofs}
The perpendicular shifts are given by, 
\begin{subequations}
\begin{align}
    \Delta E_\perp &= \left(\beta - 1\right)  \xi_a\dot{\shift}_\perp^a - \fr{1}{2}\xi_{a;b}\mf{G}^{ab} \label{eqn: perp-shifts-E}\\
    \Delta {L}_{\perp} &= \left(1-\beta\right)\eta_a\dot{\shift}_\perp^a +\fr{1}{2}\eta_{a;b}\mf{G}^{ab} \label{eqn: perp-shifts-L}\\
    \Delta \GC_{0\perp} &= 2u^a \mf{G}^{bc}L_{abc} - 2\left(\beta-1\right) K_{ab}v^a\dot{\shift}_\perp^b~, \phantom{\frac12} \label{eqn: perp-shifts-CC}\\
    \Delta N_\perp &= -2\left(\beta-1\right) v_a\dot{\shift}_\perp^a~. \label{eqn: perp-shifts-norm}
\end{align}    
\end{subequations}

Let us start with the energy shift. Recall that $\mf{G}$ is orthogonal to both $f^\star_{ab}$ and $f_{ab}$. The Papapetrou field $\xi_{a;b}$ lies in the plane spanned by $f_{ab}$ and $f^\star_{ab}$. This is made precise (for Type D spacetimes) with the identity \eqref{eq:KY-dxi-typeD}
\begin{equation*}
    \xi_{a;b} = \left(\mathrm{Re}[\Psi_2] + \fr{R}{12}\right) f^\star_{ab} - \mathrm{Im}\left[{\Psi_2}\right]f_{ab}~,
\end{equation*}
where $R$ is the Ricci scalar and $\Psi_2$ is the only non-vanishing Weyl scalar in a Type D spacetime. From this decomposition, we can immediately see that the term $\xi_{ab}\mf{G}^{ab}$ must vanish. However, we have already shown in \eqref{eqn: derdxperp} that $\dot{\shift}_\perp$ is also orthogonal to $\xi$. Therefore, the perpendicular energy shift vanishes identically, for any choice of $\beta$. 

For the angular momentum, consider the second term of $\Delta {L}_\perp$. We first simplify $\mf{G}^{ab}$ by rewriting \eqref{eqn: perp-spin-tensor-fancy-G} entirely in terms of $f_{ab}$, $v^a$ and $e_\perp^a$
\begin{equation}
    \mf{G}_{ab} = -\fr{2 \ui}{\sqrt{\GC_0}}\left(v_{[a}f_{b]c}e_\perp^c - e^\perp_{[a}f_{b]c}v^c\right)~.
\end{equation}
Due to the antisymmetry, the contraction with $\eta_{a;b}$ simplifies as 
\begin{align}
    \fr{1}{2}\eta_{a;b}\mf{G}^{ab} &= -\fr{\ui}{\sqrt{\GC_0}}\eta_{a;b}\left(v^a f^b_{\ph c} e_\perp^c - e_\perp^af^b_{\ph c}v^c\right)= \fr{2\ui}{\sqrt{\GC_0}}\eta_{a;b}f^a_{\ph c}v^{[b}e_\perp^{c]}\,.
\end{align}
The gradient $\eta_{a;b}$ can be written down in terms of the KY and the principal tensor as in \eqref{eq:KY-deta-typeD}
\begin{equation}
    \eta_{a;b} = 2\xi^c f^\star_{c[b}\xi_{a]} - \xi^2 f^\star_{ab} + \mf{Z} \left(\rl{\Psi_2} + \fr{R}{12}\right) f_{ab} - \img{\Psi_2}K_a^{\ph c} f_{cb}~.
\end{equation}
From the orthogonality properties of $\mf{G}$, we see that only the first term in the above expansion of $\eta_{a;b}$ will contribute to this contraction. This non-vanishing contribution is 
\begin{equation*}
    2\eta_{a;b}f^a_{\ph c}v^{[b}e_\perp^{c]} = -2\xi^b \xi^c f^\star_{b a}f_{cd}v^{[d}e_\perp^{a]} = 2\xi^c \xi^d f^\star_{c[a}f_{b]d}e_\perp^{a}v^b~,
\end{equation*}
where we have transferred the antisymmetry in the $d,a$ index pair downstairs. To proceed further, we perform some index gymnastics and rewrite our expressions using the Kronecker delta. 
\begin{align*}
    2\xi^c\xi^d f^\star_{c[a}f_{b]d}e_\perp^{a}v^b &= \xi^c\xi^d\left(f^\star_{ca}f_{bd}-f^\star_{cb}f_{ad}\right)e_\perp^{a}v^b~, \\
    &= -\xi^c\xi^df_{de}\left(\delta^e_c f^\star_{ab} - \delta^e_a f^\star_{cb} + \delta^e_b f^\star_{ca}\right)e_\perp^{a}v^b~, \\
    &= -\xi^c\xi^d f_{de}\varepsilon_{lcab}f^{le}e_\perp^{a}v^b = K^c_{d}\xi^d f_{ca;b} e_\perp^{a} v^b = \eta^c f_{ca;b}e_\perp^{a}v^b~, \\
    &= \ui \sqrt{\GC_0}\eta_a\dot{\shift}_\perp^a~, 
\end{align*}
where, in the last equation we use \eqref{eq:KY-xi}, and in going from the second to third equality we use $\varepsilon_{dabc}f^{de} = \delta^e_a f^\star_{bc} - \delta^e_b f^\star_{ac} + \delta^e_c f^\star_{ab}$. We have thus shown that 
\begin{equation}
    \fr{1}{2}\eta_{a;b}\mf{G}^{ab} = - \eta_a\dot{\shift}_\perp^a.
\end{equation}
This must mean that in the expression for $\Delta {L}_\perp$, the only surviving term is linear in $\beta$, 
\begin{equation}
    \Delta {L}_\perp = -\beta \eta_a\dot{\shift}_\perp^a\,.
\end{equation}

Moving on to the Carter constant with $L_{abc}= - 2 f^d_{\ph [a}f_{c]b;d}$, in \eqref{eqn: perp-shifts-CC} notice that this object is contracted with $\mf{G}^{bc}$. Since $\mf{G}$ is antisymmetric, only the antisymmetric part of $L$ will contribute to this contraction. This is given by, 
\begin{equation}
    L_{a[bc]} = f^d_{\ph a} f_{bc;d} + f^d_{\ph [b}f_{c]a;d}~.
\end{equation}
Evaluating the contraction with only the antisymmetric part and using the expression in \eqref{eqn: perp-spin-tensor-fancy-G} one can show that the first term simplifies to
\begin{align*}
     v^a \mf{G}^{bc}L_{abc} 
    &= -\fr{\ui}{\sqrt{\GC_0}}K_{ab}v^a e_\perp^d f^b_{\ph c;d} v^c.
\end{align*}
Therefore, the full expression for the perpendicular shift in the Carter constant is 
\begin{align}
    \fr{1}{2}\Delta \GC_{0\perp} &= -\fr{\ui}{\sqrt{\GC_0}}K_{ab}v^a e_\perp^d f^b_{\ph c;d} v^c + \fr{\ui}{\sqrt{\GC_0}}\left(\beta - 1\right)K_{ab}v^a f^b_{\ph c;d}e_\perp^c v^d  \nonumber \\ 
    &= \beta \fr{\ui}{\sqrt{\GC_0}}K_{ab}v^a f^b_{\ph c;d}e_\perp^c v^d  = -\beta K_{ab}v^a \dot{\shift}_\perp^b~.
\end{align}

Finally the four velocity normalisation has the perpendicular shifts given by $\Delta N_\perp= -2\left(\beta-1\right)\dot{\shift}_\perp^a v_a = 0$ trivially from \eqref{eqn: derdxperp}.

\subsection{Parallel shifts for the constants of motion}
\label{app:par-proofs}

We have expressions for the parallel shifts to the physical quantities below: 
\begin{subequations}
\begin{align}
    \Delta E &=\left(\alpha -1\right)\xi_a\dot{\shift}_3^a - \fr{1}{2}\xi_{a;b}\mf{F}^{ab} \label{eqn: parallel-shifts-E},\\ 
    \Delta {L} &=  \fr{1}{2}\eta_{a;b}\mf{F}^{ab} \label{eqn: parallel-shifts-L}-\left(\alpha -1\right)\eta_a\dot{\shift}_3^a,\\ 
    \Delta \GC_0 &=2\left(v^a\mf{F}^{bc}L_{abc}-\left(\alpha-1\right)K_{ab}v^a \dot{\shift}_3^b\right), \label{eqn: parallel-shifts-CC} \phantom{\fr{1}{2}}\\
    \Delta N&= 2\left[\left(\alpha -1\right)v_a\dot{\shift}_3^a\right], \label{eqn: parallel-shifts-norm} 
\end{align}
\end{subequations}
with the unknown coefficient $\alpha$. Once again, we note that $\Delta N$ vanishes for all choices of $\alpha$. 

For the Carter constant, one can directly evaluate the parallel shift by noting that 
\begin{equation}
    f_\star^{bc}L_{abc} = 2 \eta_a~.
\end{equation}

Thus, the Carter constant shift is evaluated to be
\begin{equation}
    \Delta \GC_0 = 2\left[(1-\alpha)E\,\GC_0 + (3-\alpha){L} \right] \GC_0^{-1/2},
\end{equation}
which is conserved for any constant value of $\alpha$. 

To simplify the parallel energy shift, we first expand out the derivative of the vector $\shift_3$ using \eqref{eq:derdx3}. We find that the shift contains some terms which are conserved by construction i.e. 
\begin{equation*}
    \sqrt{\GC_0} \Delta E = -\left(\alpha -1\right)E^2 - \left(\alpha -1\right)\xi^2 + \fr{1}{2}f_\star^{ab}\xi_{a;b}~.
\end{equation*}
When taking time derivatives, the first term vanishes since $E$ is the MPTD conserved energy. For the remaining terms, our strategy will be to use Kostant's formula (see C.3.6. in \cite{Wald}) for the second derivative of the Killing vector $\xi^a$ and to explicitly evaluate the derivative term coming from $f^\star_{ab}$ using the defining property of the principal tensor. 
Denoting the derivatives with overdots, we see that the derivative of $f^\star_{ab}$ simplifies to  
\begin{equation*}
    \dot{f}_\star^{ab}\xi_{a;b} = v^c \left(\delta^a_c \xi^b - \delta^b_c \xi^a\right)\xi_{a;b} = 2v^{[a}\xi^{b]}\xi_{a;b} = -2\xi_a \dot{\xi}^a~. 
\end{equation*}
For the first term, we make use of Kostant's formula relating the second covariant derivative of the Killing vector $\xi$ with the Riemann tensor, and then decompose the Riemann tensor into the Weyl tensor and Ricci tensor pieces \cite{Stephani:2003tm,Ra.Iso.Dru.IntegO2.26}. 
\begin{equation}
    v^c f_\star^{ab}\xi_{a;bc} 
    = v^c \left(C_{cdab}f_\star^{ab} - \tfrac{1}{3}Rf^\star_{cd} + 2f^\star_{a[d}R_{\ph c]}^a\right)\xi^d 
    = 4 \xi_a \dot{\xi}^a + 2f^\star_{a[d}R_{\ph c]}^a v^c \xi^d,
\end{equation}
where we have used \eqref{eq:KY-dxi} in the second equality the result. Putting all of this together, we see that many of the terms combine: 
\begin{equation}
    v^c \nabla_c \left(\Delta E\right) = \fr{1}{\sqrt{\GC_0}}\left[\left(3-2\alpha\right)\xi_a \dot{\xi}^a +  f^\star_{a[d}R_{\ph c]}^{a} v^c \xi^d\right]~.
\end{equation}
The object $f^\star_{ad}R_{~c}^{a}$ is automatically antisymmetric in the indices $(d,c)$ from the curvature constraints \eqref{eq:KY-Ricci-constraints}, so we will drop the notation henceforth. A similar analysis applies to the angular momentum which gives us the results \eqref{eqn:dE-dL-cons}.

\subsection{Covariant derivation of the Lorentz-force equation}
\label{app:cov-proof}

In this appendix we derive covariantly Eq.~\eqref{eq:Lorentz-nu}. Every term carrying a factor $s_\parallel$ or $s_\perp$ is already $O(s)$, so it is evaluated on the background geodesic, where $u^a\simeq v^a\simeq\nu^a$. We write
\begin{equation}
    \dot\xi^a:=\nu^b\nabla_b\xi^a\,,\quad F^a:=f^a{}_g\,\nu^g=-\sqrt{\GC_0}\,e_3^a\,,\quad w^a:=\rl{e_\perp^a}\,,\quad y^a:=\img{e_\perp^a}\,,
\end{equation}
where the second equality is Eq.~\eqref{eqn: Marck-tetrad-e3}. 
Extending $\nu^a$ to a vector field by holding the first integrals \eqref{eqn:Gen-Shifted-Constants} fixed, $\nu^a$ is invariant under the isometry generated by $\xi^a$, since the first integrals, the metric, and the Killing--Yano tensor all are. Hence $\mathcal{L}_\xi\nu^a=0$, i.e., $\xi^b\nabla_b\nu^a=\nu^b\nabla_b\xi^a=\dot\xi^a$, as can also be checked explicitly from Eq.~\eqref{eq:nu-sep}.

The proof has two parts: Part~1 amounts to substituting relations established in the main text and ends with the equation of motion for $\nu^a$ up to a perpendicular source denoted $Q^a$; and Part~2 shows that $Q^a=0$, essentially implied by the algebraic Bianchi identity.

\emph{Part 1. Equation of motion for $\nu^a$}

Differentiating the relation between the tangents, Eq.~\eqref{eqn: two-tangents}, and using the geodesic-deviation equation for the separation $\shift^a$ between $x$ and $\tilde x$, we find
\begin{equation}\label{eq:A-kin}
    \frac{D\tilde u^a}{\ud\pt}=\frac{Du^a}{\ud\pt}+\frac{D^2\shift^a}{\ud\pt^2}+R^a{}_{bcd}\,\nu^b\shift^c\nu^d\,.
\end{equation}

The MPTD equations \eqref{EElin} and \eqref{momvellin} and the spin-shift
 $s^{bc}=\tilde s^{bc}-\nu^b\shift^c+\nu^c\shift^b$ (cf. Eq.~\eqref{eqn: two-tangents}) turn \eqref{eq:A-kin} into
\begin{equation}\label{eq:A-force}
    \frac{D\tilde u^a}{\ud\pt}=R^a{}_{bcd}\,\tilde s^{bc}\,\nu^d+\frac{D^2\shift^a}{\ud\pt^2}\,,
\end{equation}
where we used the symmetries of the Riemann tensor $R_{abcd}\nu^c\nu^d=0$ and the equality between the various velocities at leading order. 

We then insert into equation \eqref{eq:A-force} several things. First, Eq.~\eqref{eq:Marck-decomp-spin-tensor}, $\tilde s^{bc}=s_\parallel\mf{F}^{bc}+s_\perp\rl{\mf{G}^{bc}}$, and the expansion $\shift^a=s_\parallel\shift_3^a+s_\perp\rl{\shift_\perp^a}$ into \eqref{eq:A-force}. The spin components $s_\parallel$ and $s_\perp$ are constant, so they factor out of the derivatives. Second, from Eq.~\eqref{eq:nu-ansatz} (with $\beta=0$), we get
\begin{equation}
    \frac{D\nu^a}{\ud\pt}=\frac{D\tilde u^a}{\ud\pt}-\alpha\,s_\parallel\,\ddot\shift_3^a. 
\end{equation}
The second derivatives of the shifts $\sigma_3^a$ and $\sigma_\perp^a$ are then given in Eq.~\eqref{ddotshifts}. Third, using Eqs.~\eqref{eqn: spin-tensor-split-par}, \eqref{eqn: deriving-dx3}, \eqref{eq:KY-dual} and the definition Eq.~\eqref{eq:def-F} give altogether
\begin{equation}
    s_\parallel R_{abcd}\,\mf{F}^{bc}\nu^d=-2q\,\dot\xi_a+q\,F_{ad}\nu^d\,.
\end{equation}
where the definition of the effective charge was used. Fourth, by definition,  $D\nu^a/\ud\pt=\tilde u^b\nabla_b\nu^a=\nu^b\nabla_b\nu^a+\alpha s_\parallel\dot\shift_3^b\nabla_b\nu^a$. With Eq.~\eqref{eq:derdx3}, we find that $D\nu^a/\ud\pt=\nu^b\nabla_b\nu^a-\alpha\,q\,\dot\xi^a$, to leading order. Collecting the outputs of these four steps finally gives 
\begin{equation}\label{eq:A-Q}
    \nu^b\nabla_b\nu^a=q\,F^a_{\ph b}\,\nu^b+\left(2\alpha-3\right)q\,\dot\xi^a+s_\perp \Bigl( \underbrace{R^a_{\ph bcd}\,\rl{\mf{G}^{bc}}\nu^d+ \GC_0^{-1/2}\,\varepsilon^a_{\ph bcd}\,y^b\dot\xi^c\nu^d}_{Q^a}\Bigr)\,.
\end{equation}
For $\alpha=3/2$ the $\dot\xi_a$ term cancels, and it remains to show $Q^a=0$.

\emph{Part 2. Proof that the perpendicular source $Q^a$ vanishes}

From Eq.~\eqref{eqn: spin-tensor-split-perp} and Eq.~\eqref{eqn: deriving-dxperp}, $\rl{\shift_\perp^c}=f^c{}_g\,y^g/\sqrt{\GC_0}$. The orthonormality and orientation of the Marck tetrad, Eq.~\eqref{eq:def-Marck-tetrad}, give $\varepsilon^{bc}{}_{ef}\nu^ew^f=(F^by^c-F^cy^b)/\sqrt{\GC_0}$. Therefore, using the symmetries of the Riemann tensor, one finds the first term in $Q^a$ as
\begin{equation}\label{eq:A-G}
    \sqrt{\GC_0}\,R_{abcd}  \, \rl{\mf{G}^{bc}}\nu^d=R_{abcd}\,\nu^b f^c{}_gy^g\,\nu^d+R_{abcd}\,F^by^c\,\nu^d-R_{abcd}\,F^cy^b\,\nu^d\,.
\end{equation}

For the second term in $Q^a$, cf. \eqref{eq:A-Q}, we start from the Ricci identity for $f_{ab}$ \cite{Ra.Iso.IntegO1.26}
\begin{equation}
   2\nabla_{[a}\nabla_{b]} f_{cd}=R_{abc}{}^{e}f_{ed}+R_{abd}{}^{e}f_{ce}\,.
\end{equation}
We contract it with $\nu^b\nu^cy^d$. On the left, we use Eq.~\eqref{eq:KY-xi} and the antisymmetry of the Levi-Civita tensor. On the right, we use $f_c{}^e\nu^c=-F^e$. This gives
\begin{equation}\label{eq:A-KYint}
    \varepsilon_{acde}\,\nu^cy^d\dot\xi^e=R_{abce}\,\nu^b\nu^c\,f^e{}_gy^g-R_{abde}\,\nu^by^d\,F^e\,, 
\end{equation}
which is, up to a cyclic permutation of indices, the second term in $Q^a$, cf. \eqref{eq:A-Q}. Combining \eqref{eq:A-G} with \eqref{eq:A-KYint}, we obtain 
\begin{equation}
    \sqrt{\GC_0}\,Q_a=-(R_{abcd}\,\nu^by^cF^d+R_{abcd}\,y^bF^c\nu^d+R_{abcd}\,F^b\nu^cy^d)=-3\,R_{a[bcd]}\nu^by^cF^d\,,
\end{equation}
which vanishes owing to the first Bianchi identity. Since $Q^a=0$, Eq.~\eqref{eq:A-Q} reduces to
\begin{equation}
    \nu^b\nabla_b\nu^a=q\,F^a{}_b\,\nu^b+\left(2\alpha-3\right)q\,\dot\xi^a + O(s^2)\,,
\end{equation}
which is Eq.~\eqref{eq:Lorentz-nu} for $\alpha=3/2$. 
This is precisely the value that removes the first term in Eqs.~\eqref{eqn:dE-dL-cons}, so the conserved quantities and the Lorentz-force form single out the same choice.

\subsection{Proofs of identities}
\label{app:ids}

In this appendix, we prove identities \eqref{eq:KY-dxi-typeD}, \eqref{eq:KY-deta-typeD} and \eqref{eqn:pot-origins}.

\emph{Proving \eqref{eq:KY-dxi-typeD}}

Starting with \eqref{eq:KY-dxi}, we expand the Riemann tensor in terms of the Weyl tensor and Ricci tensor pieces. The Weyl tensor can then be expanded in terms of the Weyl scalars in a (complex) self-dual bivector basis $\{X, Y, Z\}$ (see \cite{Stephani:2003tm,Ra.Iso.Dru.IntegO2.26})
\begin{equation*}
    C_{abcd} = \Psi_2\left(Z_{ab}Z_{cd} - X_{ab}Y_{cd} - Y_{ab}X_{cd}\right) + \text{c.c.}~, 
\end{equation*}
where c.c. denotes complex conjugation. Since the spacetime is of Petrov type D (as implied by the non-degenerate Killing–Yano tensor), $\Psi_2$ is the only non-vanishing Weyl scalar. The KY and the principal tensor are then real and imaginary parts of the basis bivector $Z_{ab}$. Then the expression simplifies significantly when we express $Z$ and $\bar{Z}$ in terms of $f$ and $f^\star$ and compute the required contraction in \eqref{eq:KY-dxi} (note that $\{X,Y,Z\}$ forms an orthogonal basis). We thus obtain 
\begin{equation}
    \xi_{a;b} = \left(\rl{\Psi_2} + \fr{R}{12}\right) f^\star_{ab} - \img{\Psi_2} f_{ab}~.
\end{equation}

\emph{Proving \eqref{eq:KY-deta-typeD}}

Recall that $\eta_a = K_{ab}\xi^b$. Then $\eta_{a;b} = K_{ac;b}\xi^c + K_{a}^c \xi_{c;b}$. We now simplify each term individually
\begin{align*}
K_a^c \xi_{c;b} &= 
f_{ad} f^{cd}\left\{ \left(\rl{\Psi_2} + \fr{R}{12}\right) f^\star_{cb} - \img{\Psi_2} f_{cb}\right\} \cr 
&= \mf{Z}\left(\rl{\Psi_2} + \fr{R}{12}\right) f_{ab} - \img{\Psi_2}K_a^c f_{cb}~.
\end{align*}
This gives us the first two terms of the identity. For the last two terms, we have 
\begin{align*}
K_{ac;b}\xi^c = \xi^c \left(f_{ad;b}f_{c}~^{d}+f_{cd;b}f_{a}~^{d}\right)~.
\end{align*}
We know that the derivative of the KY is proportional to the Hodge dual of the primary KV \eqref{eq:KY-xi}. This means the second term vanishes i.e. 
$f_{cd;b}\,\xi^c = -\varepsilon_{bcde}\,\xi^c\xi^e = 0$

With the only surviving term, we proceed similarly, by using \eqref{eq:KY-xi} and rewriting $f = -f^{\star\star}$. Then, 
\begin{align*}
    K_{ac;b}\xi^c &= -2\xi^c f^\star_{c[a}\xi_{b]} - \xi^2 f^\star_{ab}~.
\end{align*}
Putting the two terms together, proves the identity. 

\emph{Proving \eqref{eqn:pot-origins}}

Let us start by directly taking derivatives of $f^{\star ab}\xi_{a;b}$, 
\begin{align*}
    \nabla_c\left(f^{\star ab}\xi_{a;b}\right) &= f^{\star ab}~_{;c}\xi_{a;b} + f^{\star ab}\xi_{a;bc}~, \\
    &= \left(\delta^a_c\xi^b-\delta^b_c\xi^a\right)\xi_{a;b} + \left(C_{abcd}-\fr{R}{3} g_{a[c}g_{d]b} + g_{a[c}R_{d]b} - g_{b[c}{R}_{d]a}\right)f^{\star ab}\xi^d~, \\
    &= 2\xi^b\xi_{c;b} + \left(C_{abcd}f^{\star ab} - \fr{R}{3}f^{\star}_{cd}\right)\xi^d + \left(f^{\star ab}R_{c}~^a-f^{\star ac} R_{b}~^a\right)\xi^b~, \\
    &= 2\xi^b\xi_{c;b} - 4\xi_{c;d}\xi^d + 2 f^{\star}_{ab}R^a~_{c}\xi^b~,
\end{align*}
where we have used Kostant's formula \cite{Wald}, the curvature constraints \eqref{eq:KY-Ricci-constraints}, and the decomposition of the Riemann tensor into Weyl and Ricci pieces. 
Since $\nabla_c(\xi^2) = 2\xi^a\xi_{a;c} = -2\xi^b\xi_{c;b}$, the first two terms combine with it, and we obtain $\nabla_c\Psi_\xi = 2f^\star_{ab}R^a{}_c\xi^b$. The identity for
$\Psi_\eta$ follows in the same way, using $\eta_{a;b}\xi^b = \xi_{a;b}\eta^b$.

\end{document}